\documentclass[preprint,twocolumn,preprintnumbers,aps,nofootinbib,tightenlines,floatfix,groupedaddress]{revtex4-1}

\def\Nslash{{ {\cal N}\hskip-0.55em /}}

\usepackage{hyperref}
\usepackage{mathtools}
\usepackage{amsmath,amssymb}
\usepackage{cleveref}
\usepackage{graphicx}
\usepackage{xcolor}
\usepackage[utf8]{inputenc}
\usepackage{bm}
\usepackage{csquotes}
\usepackage[section]{placeins}
\usepackage{csquotes}
\usepackage{scalerel}
\usepackage{comment}
\usepackage[export]{adjustbox}
\usepackage{booktabs}

\allowdisplaybreaks

\newcommand{\lsim}{\raisebox{-0.7ex}{$\stackrel{\textstyle <}{\sim}$ }}

\newcount\hour \newcount\hourminute \newcount\minute
\hour=\time \divide \hour by 60
\hourminute=\hour \multiply \hourminute by 60
\minute=\time \advance \minute by -\hourminute
\newcommand{\mydate}{\ \today \ - \number\hour :\number\minute}

\makeatletter
\renewcommand*\env@matrix[1][*\c@MaxMatrixCols c]{%
  \hskip -\arraycolsep
  \let\@ifnextchar\new@ifnextchar
  \array{#1}}
\makeatother

\begin{document}

\title{Geometric Quantum Information Structure in Quantum Fields and their Lattice Simulation
}

\author{Natalie Klco}
\email{klcon@uw.edu}
\thanks{Affiliation after September 2020: Institute for Quantum Information and Matter and Walter Burke Institute for Theoretical Physics, California Institute of Technology, Pasadena CA 91125, USA}
\affiliation{Institute for Nuclear Theory, University of Washington, Seattle, WA 98195-1550, USA}
\author{Martin J.~Savage}
\email{mjs5@uw.edu}
\affiliation{Institute for Nuclear Theory, University of Washington, Seattle, WA 98195-1550, USA}

\date{\mydate}
\preprint{INT-PUB-20-031}

\begin{figure}[!t]
 \vspace{-1.5cm} \leftline{
 	\includegraphics[width=0.12\textwidth]{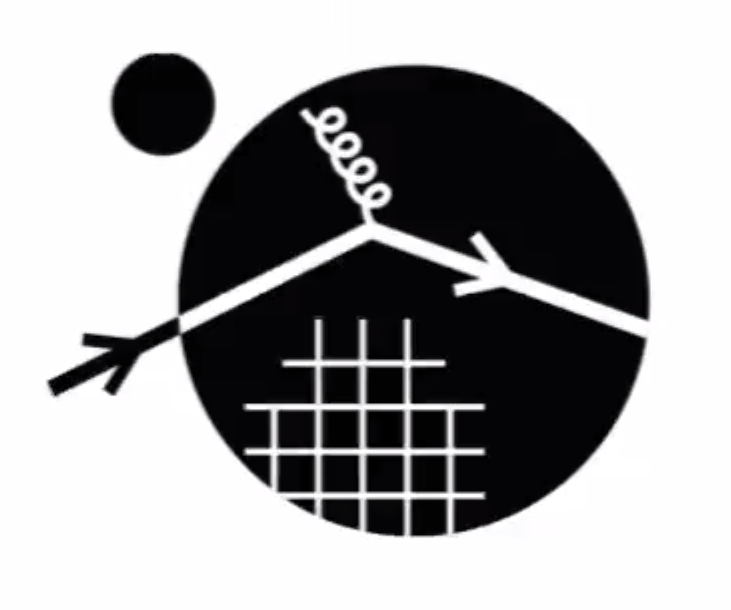}}
\end{figure}

\begin{abstract}
An upper limit to distillable entanglement between two disconnected regions of massless non-interacting scalar field theory has an exponential decay defined by a geometric decay constant. When regulated at short distances with a spatial lattice, this entanglement abruptly vanishes
beyond a dimensionless separation, defining a negativity sphere.
In two spatial dimensions, we determine this geometric decay constant between a pair of disks and the growth of the negativity sphere toward the continuum through a series of lattice calculations.
Making the connection to quantum field theories in three-spatial dimensions,
assuming such quantum information scales appear also in quantum chromodynamics (QCD),
a new relative scale may be present in
effective field theories describing the low-energy dynamics of nucleons and nuclei.
We highlight potential impacts of the distillable entanglement structure on effective field theories, lattice QCD calculations and future quantum simulations.
\end{abstract}

\maketitle

\section{Introduction}
\noindent
It is well known that the vacuum state of quantum fields exhibits entanglement between spatially separated regions~\cite{Reeh1961,summers1985vacuum,summers1987bell1,summers1987bell2,Witten:2018zxz}.
Techniques for extracting this entanglement to auxiliary quantum systems through harvesting and subsequent distillation
have been developed for a variety of relativistic fields, in some instances employing accelerating observers to causally
disconnect the entanglement detectors~\cite{VALENTINI1991321,Reznik:2002fz,Reznik:2003mnx,Salton_2015,Pozas-Kerstjens:2015gta}.
This fundamental property of nature may prove useful in the distribution of entangled pairs through local interaction
with a background field for quantum communications, sensing, or metrology as well as in providing new perspectives
on the structure of spacetime~\cite{Ryu:2006bv,Ryu:2006ef,VerSteeg:2007xs,MartinMartinez:2012,Martin-Martinez:2015qwa,Swingle:2018rev}.

Recent progress in quantum information has inspired increased consideration of entanglement in high-energy physics and nuclear physics processes.
There have been a number of earlier studies examining the role of entanglement in dynamical processes related to
high-energy quantum chromodynamics (QCD),
such as fragmentation~\cite{Berges:2017zws,Berges:2017hne,Berges:2018cny},
heavy-ion collisions~\cite{Ho:2015rga,Kovner:2015hga,Kovner:2018rbf,Armesto:2019mna},
deep inelastic scattering~\cite{Kharzeev:2017qzs,Tu:2019ouv}
and even suggestive hints extracted from experimental data~\cite{Baker:2017wtt}.
Further, some exciting results have recently been obtained connecting entanglement to emergent
symmetries of QCD~\cite{Beane:2018oxh,Beane:2019loz} and to the structure of nuclei~\cite{GortonThesis,GortonJohnson2019a,Robin:2020aeh}.

In this work,  we calculate the geometric
constant determining the exponential component of the
decay of negativity in the two-dimensional non-interacting massless scalar field vacuum.
We further explore the structure of entanglement in the lattice-regulated field to inform the design of quantum and
classical calculations of quantum coherent observables.

The choice of scalar field is inspired by its simplicity, ubiquity, unique status of having a thoroughly examined qubit digitization~\cite{Jordan:2011ci,Jordan:2011ne,Yeter-Aydeniz:2017ubh,Klco:2018zqz,Yeter-Aydeniz:2019scalar},
and having been proven to be BQP complete~\cite{Jordan:2017lea}.
The latter of these motivating factors indicates that any efficient quantum calculation, of fields or otherwise, can be transformed with polynomial resources to a scattering process of the interacting scalar field through the manipulation of classical external sources.
As such, the entanglement structures found in the dynamical interacting scalar field are expected to be sufficient for the hardware implementation of efficient quantum computations.
Perhaps surprisingly considering the na\"ive simplicity of the massless scalar field, an analytic calculation through conformal field theory of the entanglement structure between disjoint subregions even within one spatial dimension remains elusive, complicated in part by its spectroscopic nature with respect to field correlators.
However, progress has developed through formidable analytic and high precision numerical investigations of the entanglement structure of the scalar field both in the continuum and using a spatial lattice (harmonic chains)~\cite{Srednicki:1993im,Audenaert:2002xfl,Botero_2004,kofler2006entanglement,Marcovitch:2008sxc,Lohmayer:2009sq,Calabrese:2009ez,Casini:2010kt,Calabrese:2012ew,Calabrese:2012nk,MohammadiMozaffar:2017nri,Coser_2017,DiGiulio:2019cxv}.

In the following, numerical results in the two dimensional scalar field will be used to motivate discussions related to lattice quantum chromodynamics (LQCD) calculations~\cite{Politzer:1973fx,Gross:1973id,Wilson:1974sk,Creutz:1979kf,Balian:1974ts}
and low-energy effective field theories (EFTs) describing  nuclear forces and other confining strongly-interacting theories.
Additional calculations in three dimensions are required to make quantitive predictions for previously unknown nonperturbative
systematic errors in LQCD calculations, e.g. Refs.~\cite{Beane:2012vq,Aoki:2012tk,Yamazaki:2015asa,Wagman:2017tmp},
and for potential impacts in nuclear EFTs, e.g. Refs.~\cite{Weinberg:1990rz,Weinberg:1991um,PhysRevC.49.2932,Kaplan:1998tg,Epelbaum:2019kcf}.

In order to quantify entanglement, the logarithmic negativity~\cite{Horodecki:1996nc,Peres:1996dw,Vidal:2002zz,Simon:2000zz}
between spatially separated regions of the field is calculated.
The negativity---the sum of negative eigenvalues in the partially transposed reduced density
matrix---quantifies violation of the parity symmetry in conjugate momentum space that would otherwise be exact in a tensor product state~\cite{Simon:2000zz}.
As a necessary but insufficient separability criterion,
the negativity does not capture all quantum correlations~\cite{Horodecki:1998kf},
though it does provide an upper bound~\footnote{As an upper bound, the exponentially decaying negativity calculated in the continuum massless scalar field does not preclude the possibility that the distillable entanglement of the field is zero.  Calculations harvesting entanglement from the scalar field suggest this is not the case~\cite{Pozas-Kerstjens:2015gta,Cong:2018vqx,Henderson:2020zax}.} to the distillable entanglement~\cite{Horodecki:2000,Vidal:2002zz}.

It has previously been observed that the negativity between individual oscillators of the latticized position-space scalar field vanish beyond nearest neighbor~\cite{Audenaert:2002xfl,Botero_2004,kofler2006entanglement,Marcovitch:2008sxc,Calabrese:2009ez,Calabrese:2012ew,Calabrese:2012nk,MohammadiMozaffar:2017nri,Coser_2017,Klco:2019yrb,DiGiulio:2019cxv}.
This phenomenon is analogous to ESD (entanglement sudden death or early-stage disentanglement) observed in the presence of quantum noise~\cite{PhysRevLett.93.140404,Dodd_2004,Almeida579,Yu598}, where tracing of the scalar lattice external to the regions of interest provides the mechanism of decoherence.
While the individual field operators, $\hat{\phi}$ and $\hat{\pi}$, do not produce entanglement at long distances,  individual creation/annihilation operators for position space oscillators are sensitive to entanglement at long distances.
The translation between these two bases involves a smearing in the field conjugate momentum space and points to the importance of such systematic delocalization for reproducing infrared entanglement properties through a lattice regularization.
Naturally, higher resolution of physically separated field regions through a smaller lattice spacing improves agreement with continuum symmetries~\cite{Davoudi:2012ya}.
Though expected to systematically remove artifacts associated with finite lattice spacing,
it is here found that finite negativity \emph{spheres} are not perturbatively removed
or expanded through Symanzik improvement~\cite{SYMANZIK1983187} of the lattice dispersion relation.
Supporting entangled regions of a latticized field demands a minimum information complexity of the field representation in the regions of interest, with a lattice spacing threshold only below which negativity spheres of sufficient size and accuracy are supported.

\begin{figure}[!ht]
  \includegraphics[width = 0.4\textwidth]{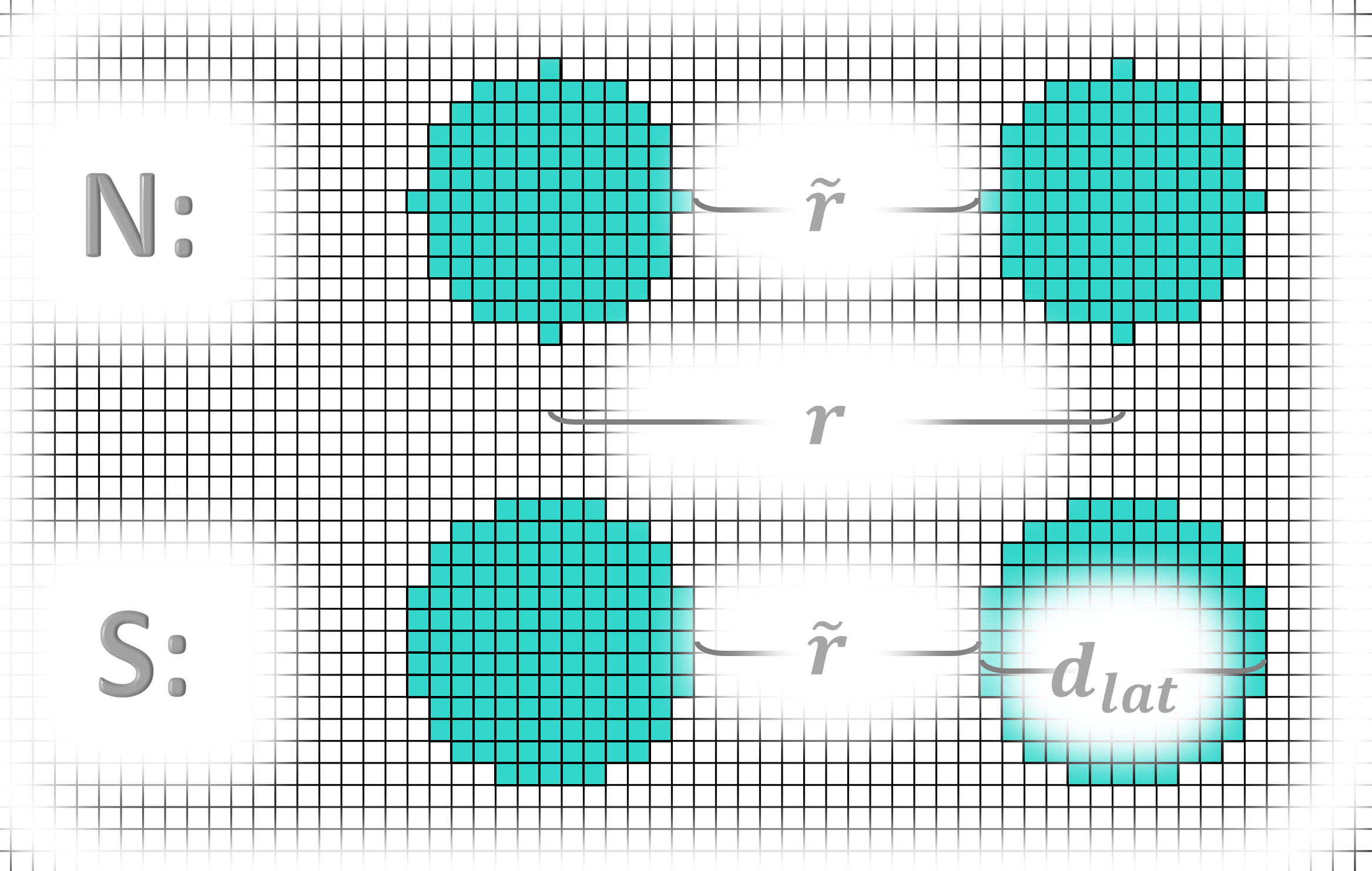}
  \caption{Two choices of pixelation producing circular regions in the continuum.  }
  \label{fig:circlediagram}
\end{figure}
Consider circular regions of a 2D massless scalar field discretized onto a lattice (see Fig.~\ref{fig:circlediagram}).
In the following, the negativity between the field degrees of freedom within such regions
with increased spatial separation is quantified.
As conformal theories are without an intrinsic length scale, allowing calculations to be
organized by relative dimensions in the continuum, all separations can be expressed in units of the region diameters e.g., $\tilde{r}/d$.
The tilde will be used to indicate a measurement of separation between the region surfaces rather than the distance between region centers.
The choice of region diameter as a reference scale allows for a natural transition
when considering the nucleon radius to set the scale for the field regions of interest in calculations of QCD.
At finite lattice spacing, the continuum limit can be approached in an arbitrary number of ways.
Two different pixelations of the regions are shown in Fig.~\ref{fig:circlediagram}.
The method labeled \enquote{$N$} begins with a central scalar field site and incorporates all sites within a specified integer radial distance.
Organizing these sites into groups by the magnitude of their vector of center displacement integers, $|\mathbf{n}|^2$~\cite{Luu:2011ep}, the $N$ boundary is defined with an integer truncation of $|\mathbf{n}|$.
The method labeled \enquote{$S$} incorporates additional $|\mathbf{n}|^2$-shells, truncating $|\mathbf{n}|$ at the next half integer.
While the $S$ boundary approaches the continuum more rapidly, the independent perspectives provided by these two trajectories toward the continuum are found to be a valuable quantifier of systematic uncertainties.
Calculating the negativity between these field regions determines
a fundamental property of the field, the distillable entanglement present within the vacuum.
A further application of this information to the operational feasibility of harvesting the present entanglement requires defining a detector structure and coupling to the studied field regions.

\begin{figure*}
  \centering
  \includegraphics[height = 0.2\textheight]{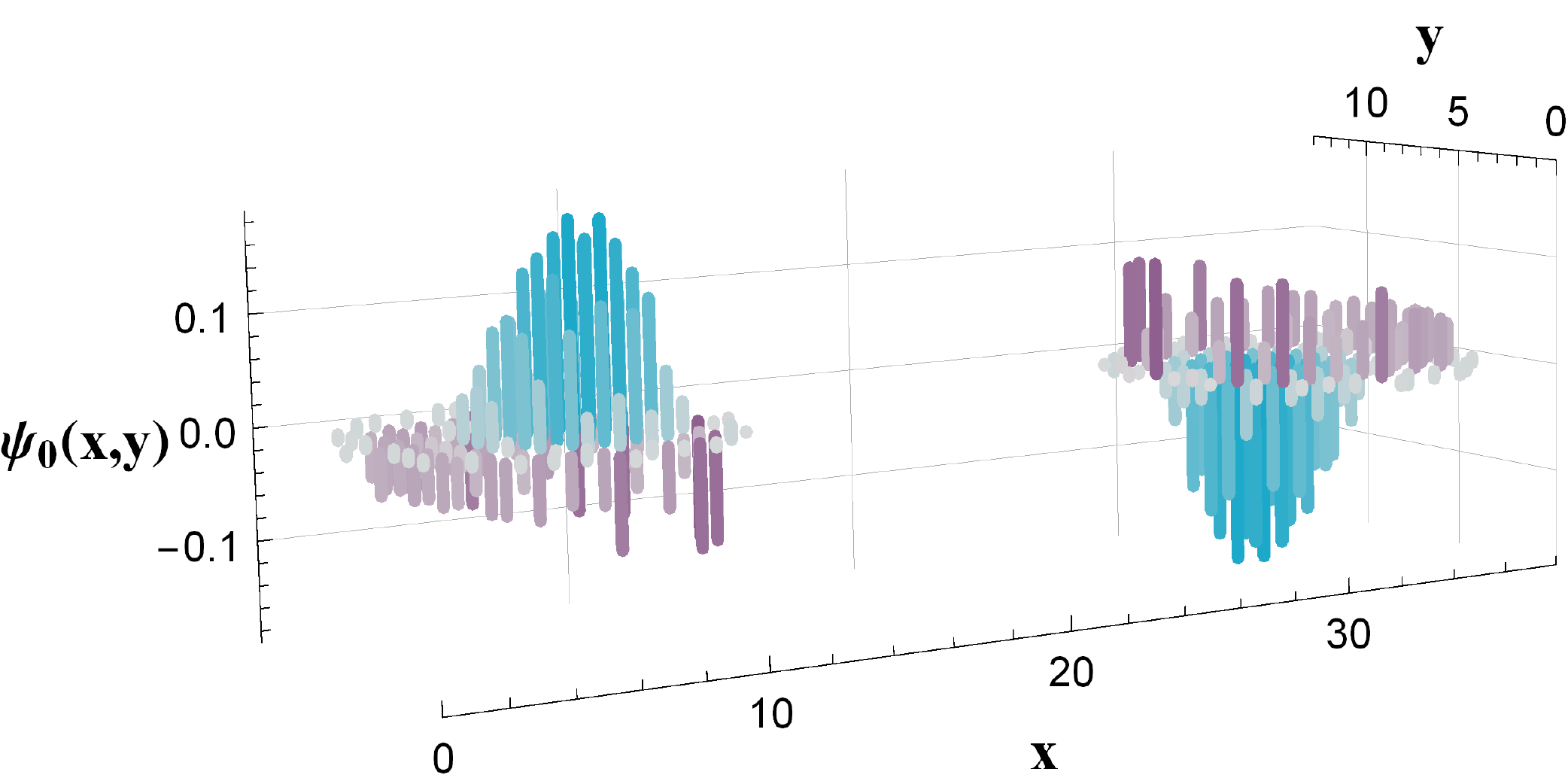} \ \
  \includegraphics[height = 0.2\textheight]{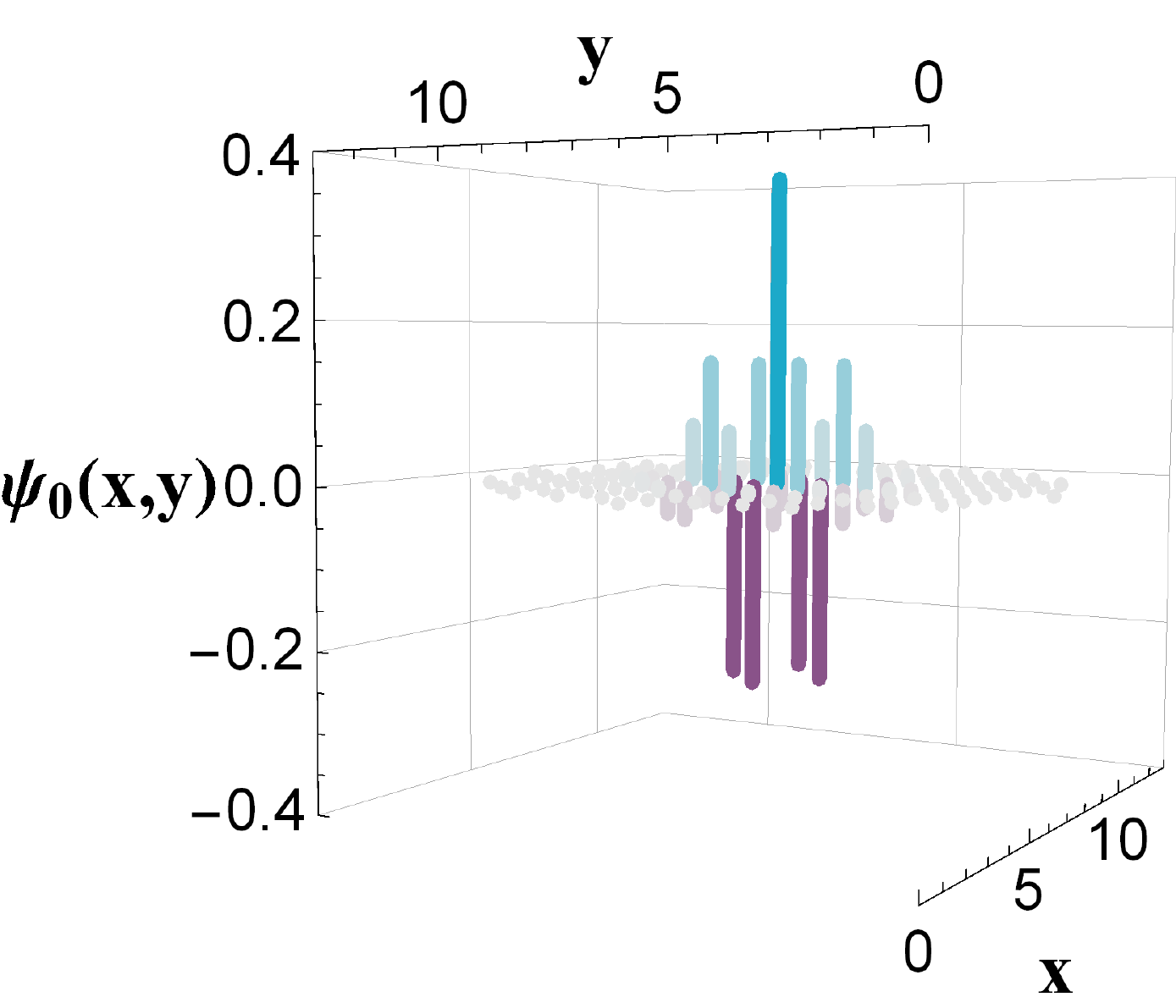}
  \caption{(left)
  The ground state wavefunction of the operator product $G H^\Gamma$
  isolated to circular regions of $d_{lat}=13$ lattice sites across with the N-type boundary
  shown in Fig.~\ref{fig:circlediagram}.
  (right) The ground state wavefunction of one isolated region calculated with $\tilde{r} = 100$, a separation beyond the negativity sphere, $\tilde{r}_{\Nslash}$.  Depicted numerical values can be found in Tables~\ref{tab:GHgammaGroundState13sep13} and \ref{tab:GHgammaGroundState13sep100} of Appendix~\ref{app:datatables}.
   }
  \label{fig:GHgammaGroundState13}
\end{figure*}
For a free non-interacting scalar field,
all observables are expressible in terms of two-point
vacuum expectation values of the field, $\hat{\phi}$, and conjugate momentum, $\hat{\pi}$, operators.
In a finite volume with spatial extent $L$ in each direction and a lattice spacing set equal to one, these two-point functions are,
\begin{multline}
  \langle \hat{\chi}(0) \hat{\chi}(\mathbf{r}) \rangle
  =
   \sum_{\mathbf{p}}
   \frac{e^{i\mathbf{p}\cdot \mathbf{r}}}{2L^D} \\ \left(m^2 + \sum_i 4 \sin^2\left(\frac{p_i}{2} \right)\right)^{\frac{\alpha}{2}}
  \ \ ,
\end{multline}
where $\alpha = 1$ for $\hat{\chi} = \hat{\pi}$, $\alpha = -1$ for $\hat{\chi} = \hat{\phi}$, and $\mathbf{r}$ is a vector of integers.
The discrete vector sum over momentum modes incorporates each spatial component taking values in the set $p_i\epsilon \frac{2 \pi}{L} \mathbb{Z}_L$ with $\mathbb{Z}_L$ a bounded set of integers between 0 and (L-1).
While it is possible to calculate entanglement properties between two separated regions through analytic
Gaussian integration of the field outside the regions to generate a reduced density operator~\cite{Srednicki:1993im,Klco:2019yrb},
it is advantageous to instead use expectation values in the thermodynamic limit, $L\rightarrow \infty$,
and properties of the symplectic spectrum to represent the calculation of entanglement with only propagators
between the two regions~\cite{Adesso:2004Extremal,Marcovitch:2008sxc}.
In this case, the logarithmic negativity can be written as,
\begin{equation}
  \mathcal{N} = -\sum_i \log_2 \min(2\sqrt{\lambda_i},1)
  \ \ \ ,
\end{equation}
where the $\lambda_i$ are eigenvalues of the matrix product $GH^\Gamma$ with
$G_{xy} = \langle \hat{\phi}(x) \hat{\phi}(y)\rangle$, $H_{xy} = \langle \hat{\pi}(x) \hat{\pi}(y)\rangle$,
and $\Gamma$ indicates the partial transposition of $H$.
Though not hermitian, the product $G H^\Gamma$ enjoys real eigenvalues associated with the symmetric positive definiteness of $G$ and $H^\Gamma$.
For interacting theories, in which higher-body correlation functions carry distinct information,
this Gaussian approximation calculated from propagators alone is expected to provide a lower bound on the logarithmic negativity of the field~\cite{wolf2006extremality}.
For this continuous variable system, the partial transposition of $H$ can be implemented with a reflection in conjugate momentum space of the second region, $\pi_2 \rightarrow - \pi_2$ ~\cite{Simon:2000zz}.
In the infinite volume limit (and  continuous momentum within the first Brillouin zone) of two-dimensional space,
the two-point correlation functions populating $G$ and $H$ can be simplified to,
\begin{multline}
  \langle \hat{\chi}(0,0) \hat{\chi}(x,y) \rangle = \\ \int_{0}^\pi \text{d} p \ \frac{ \left(6-2\cos{p}\right)^{\frac{\alpha}{2}} \cos{y p}}{2 \pi} \\ \ _3\tilde{F}_2\left( \begin{matrix} -\alpha/2,  1/2,  1 \\ 1-x,   1+x \end{matrix}; \frac{2}{3-\cos p} \right) \ \ \ ,
\end{multline}
where $_3\tilde{F}_2$ is the regularized hypergeometric function.  No infrared regulation is required in two dimensions, allowing the mass to be set to zero~\footnote{Massive theories will exhibit additional exponential suppression of the negativity controlled by the mass of the lightest particle.  The massless limit has been chosen to isolate the purely geometric contribution.}.
In this formulation,
with oscillatory integrands of increasing frequency and exponentially decreasing eigenvalues of the
product $G H^\Gamma$ in the separation, along with increasing dimensionality of $G$ and $H$ in the lattice spacing, the calculation of the negativity exhibits a sign problem.
As such, high-precision is required (typically quadruple precision or greater) for evaluations of the $G, H$ integrals and the following eigenvalue determination, limiting the granularity of achievable region pixelations.

While the point-to-point propagators can be used directly as a basis for $G$ and $H$, it is convenient to form combinations that reflect the underlying symmetry of the pixelated regions:
(1) the reflection symmetry in the plane along their separation axes and (2) the perpendicular reflection plane at the
midpoint of their separation.
This leads to a block diagonalization of $G H^\Gamma$ into the symmetry sectors of the parity operators (which remain dense matrices).
The negativity is dominated (by orders of magnitude at modest separation) by the lowest eigenvalue in the sector of (+,-) parity for reflection planes (1, 2) described above.
The wavefunction of this ground state of the product $GH^\Gamma$ is shown in the left panel of  Fig.~\ref{fig:GHgammaGroundState13} for configuration $d_{lat} = \tilde{r} = 13$.
At a separation equal to one region diameter, this configuration is within the negativity sphere.
The wavefunction shows that amplitudes experience an attractive interaction between the two regions, suggestive of a \enquote{flux-tube} between them.
With respect to the irreps of the dihedral group $D_4$, expressing a valid symmetry for individual regions and thus for a region at infinite separation, it is found that isolated contributions to the negativity are organized in the hierarchy of $\left(E, A_1, B_1, A_2, B_2\right)$ with the $E$ sector dominant at modest separations beyond the detector size.
At separations beyond the negativity spheres, this apparent flux tube is broken and the wavefunction of a region appears as in the right panel of Fig.~\ref{fig:GHgammaGroundState13}, calculated for $\tilde{r} = 100$.
The local horizontal asymmetry of each spatial region in the ground-state wavefunction  rapidly decays
and the wavefunction acquires approximately an oscillatory Gaussian envelope.
In the continuum limit at infinite separation, these regions carry zero \enquote{charge}~\footnote{The sum of the amplitudes in the wavefunction of each isolated region vanishes in the continuum.}.

\begin{figure*}[!ht]
\centering
  \includegraphics[height=0.25\textwidth]{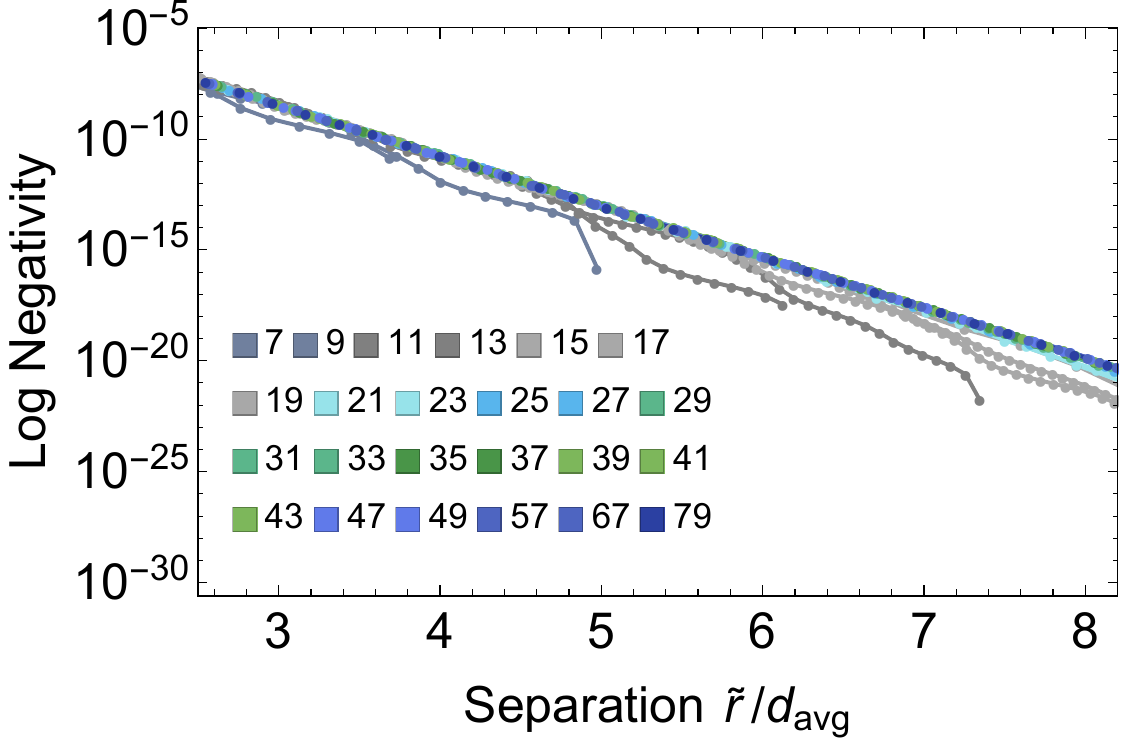}
  \includegraphics[height=0.25\textwidth]{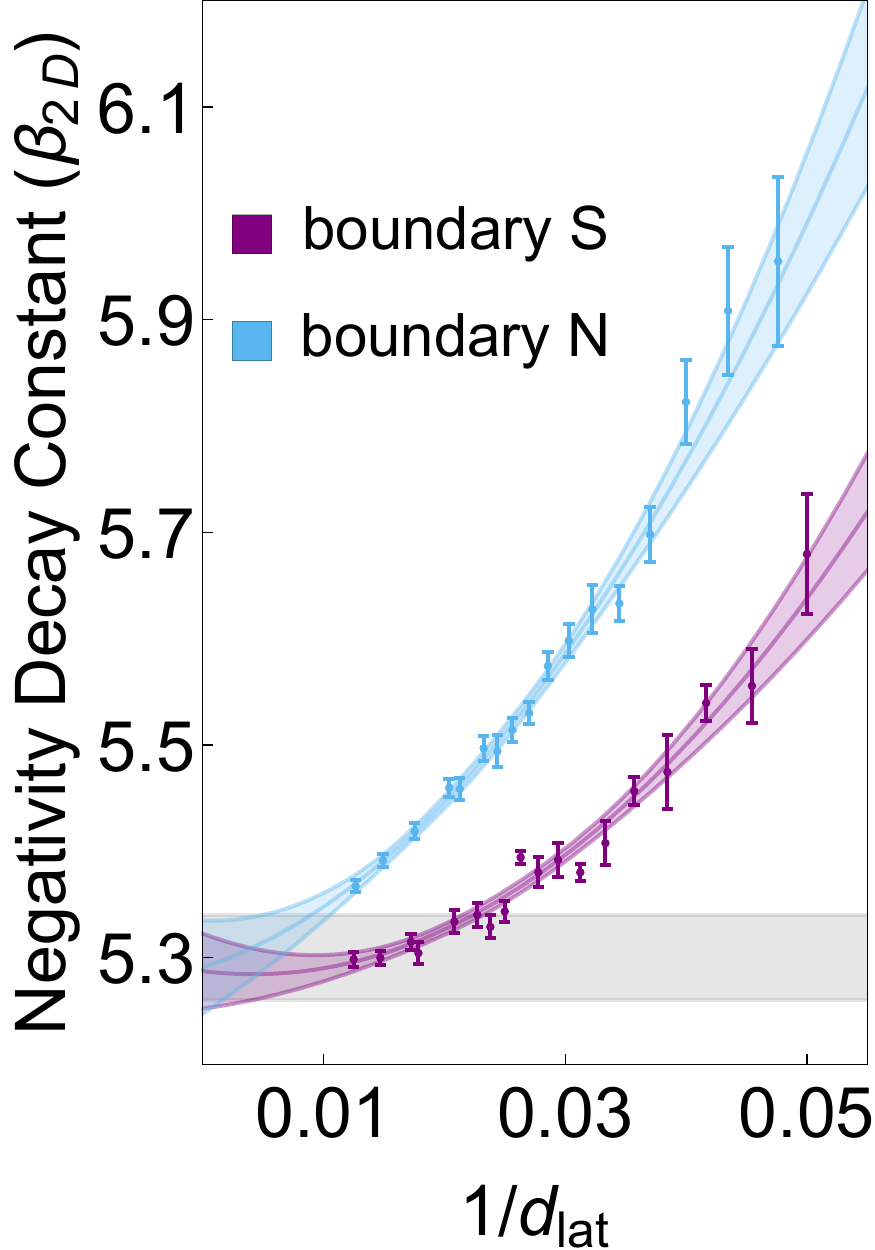}
  \includegraphics[height=0.25\textwidth]{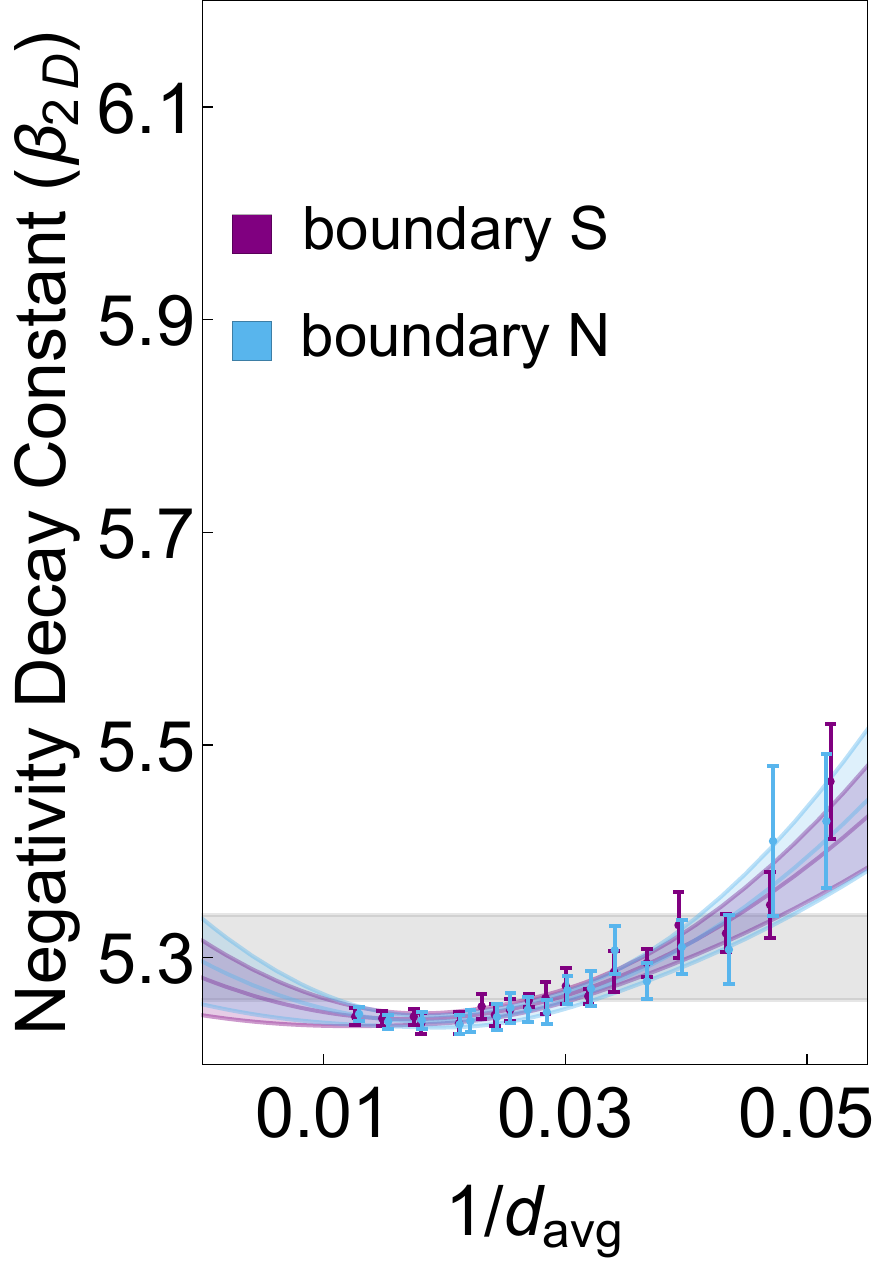}
  \includegraphics[height=0.25\textwidth]{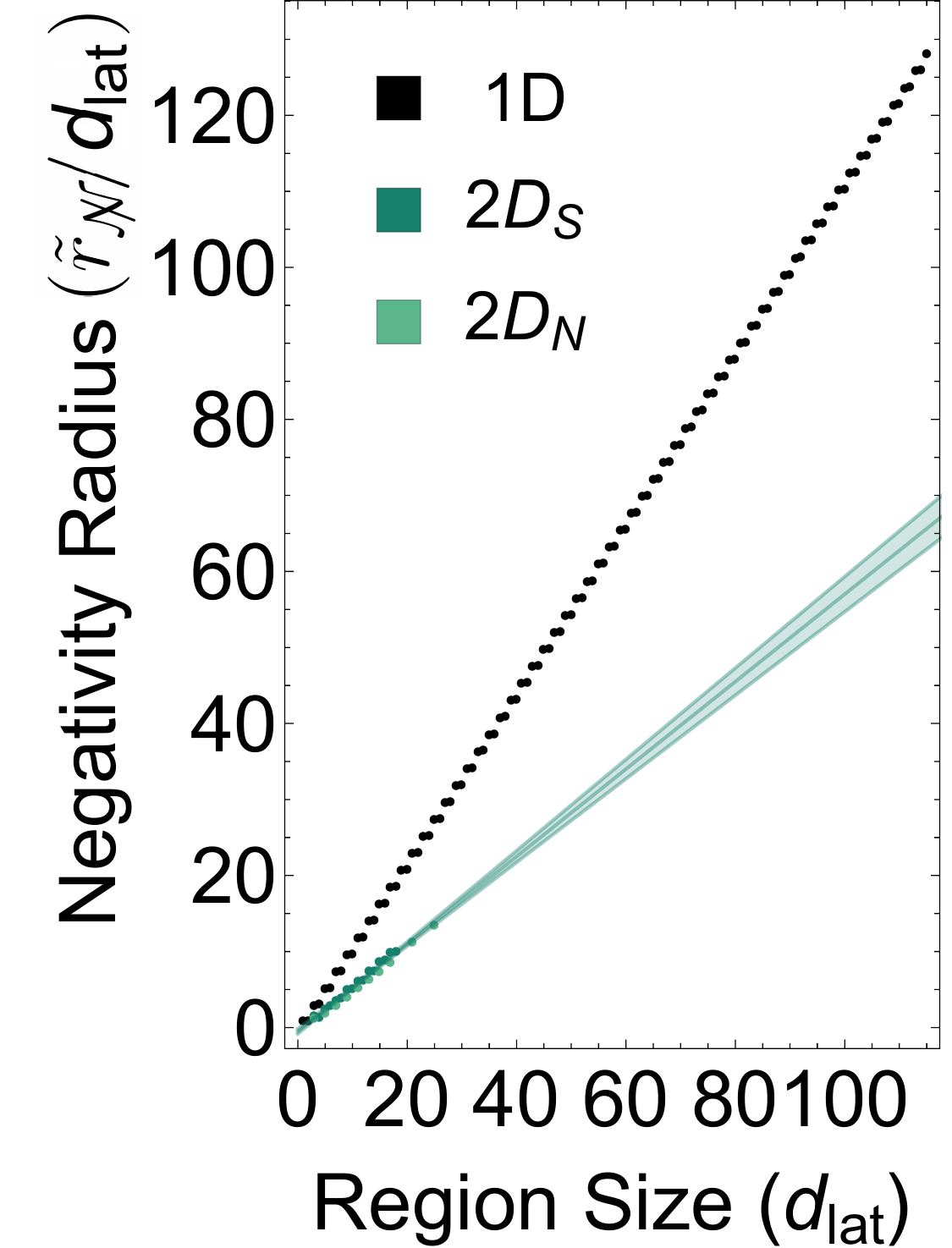}
  \caption{(left) Logarithmic negativity of two circular regions of the 2D massless non-interacting scalar field as a function of separation distance measured in units of the region diameter (see Fig.~\ref{fig:circlediagram}).
  Trajectories that end abruptly are found to exhibit zero negativity beyond a finite separation. (middle)
  Negativity decay constants, $\beta_{2D}$, extracted from the decay of logarithmic negativity as a function of the inverse region diameter (lattice spacing in units of the region diameter) extrapolated to zero with two pixelations of the circular field regions. (right) Entanglement sphere radius, $\tilde{r}_{\Nslash}$, as a function of region size in 1D and 2D. Depicted numerical values can be found in Tables~\ref{tab:2dnegativity1}-\ref{tab:2dbubbleradii} of Appendix~\ref{app:datatables}.}
  \label{fig:neg2d}
\end{figure*}
The results of computing the logarithmic negativity between these circular regions
are shown in the left panel of Fig.~\ref{fig:neg2d}.
In the continuum limit, the logarithmic negativity decay is dominated by an exponential with the
separation measured in units of the region size, $\tilde{r}/d$, where $d$ is the diameter across the region and $\tilde{r}$ is the separation between the regions,
$\mathcal{N} \sim r^\alpha\ e^{-\beta \frac{\tilde{r}}{d}} $.
Considering only the exponential behavior and not the subleading power-law~\footnote{The complete functional
form will be required for future quantitative estimates of systematic uncertainties in both classical and quantum simulations.}, this behavior  is
controlled by a pure number extrapolated in the second and third panels of Fig.~\ref{fig:neg2d} to be $\beta_{2D} = 5.29(4)$ in two dimensions.
In the second panel, the reference length scale associated with the regions, $d$, was chosen to be the largest extent of the circular region along the lattice axis (for the example regions of Fig.~\ref{fig:circlediagram}, $d_{lat} = 13$).
This classification places the $S$ and $N$ boundary structures on different trajectories toward the continuum limit of rotationally symmetric field regions.
In the third panel of Fig.~\ref{fig:neg2d}, the reference length scale associated with the field region was chosen to be an averaged diameter calculated from each point on the boundary (for the $d_{lat}=13$ example regions of Fig.~\ref{fig:circlediagram}, the $S$ boundary has $d_{avg}=12.36$ while the $N$ boundary has $d_{avg} = 11.27$).
This classification connects the trajectories of the $S$ and $N$ boundary structures.
The extrapolated entanglement mass, $\beta_{2D}$, is consistent between these two diameter definitions, though the latter is found to produce negativity within $1\%$ of the continuum value at larger lattice spacing, as expected for a continuum-inspired spatial averaging.  The resulting value of $\beta_{2D}$ is distinct from that previously calculated in one dimension, $\beta_{1D} = 2.82(3)$ ~\cite{Marcovitch:2008sxc}, indicating a more rapid decay of distillable entanglement within the massless scalar vacuum in higher dimensional space.
It is expected, and the subject of future work, that $\beta_{3D}$ will be further suppressed.

The existence of a pure number, $\beta$, in the massless non-interacting scalar field acting to exponentially suppress quantum correlations in the continuum, as would the presence of a mass scale, presents an opportunity for the appearance of an additional relative scale associated with the geometric entanglement structure in systems without conformal symmetry.
With this mechanism, the previous observation of an entanglement-based hierarchy in low-energy nuclear interactions~\cite{Beane:2018oxh}, normally predicated on the dominance of local operators, could be understood even without an explicit dimensionful parameter quantifying the entanglement.
For example, expecting $\beta_{3D}$ to be found of similar magnitude to $\beta_{2D}$, a rescaling of the radius of the nucleon set by the pion mass with a factor of $\beta$ empirically produces a mass scale of $\sim 350~{\rm MeV}$, a scale found to characterize the convergence of EFT descriptions of nucleon-nucleon interactions when accounting for the characteristic and unjustifiably large scattering lengths~\cite{Kaplan:1998tg}.
Alternatively, a na\"ive expectation that $\beta_{3D}\sim 7.7$, approximately linearly extrapolated from $\beta_{1D,2D}$, could indicate an entanglement scale associated with the pion mass to be found around the scale of chiral symmetry breaking.
As expected, the numerical results of the two-dimensional massless scalar field are not sufficient to conclusively determine the role played by the negativity decay constant, $\beta$, in the design of low-energy effective field theories of the strong interactions.
However, these preliminary considerations motivate the non-perturbative computation of $\beta_{3D}$ in both the scalar field and QCD, expected to require significant HPC computational resources.

Consistent with an understanding that the lattice representation preferentially captures local properties of the field structure and non-local properties only in the $a\rightarrow0$ limit, calculations of the negativity at finite lattice spacing non-perturbatively vanish beyond a particular radius, $\tilde{r}_{\Nslash}$.
The right panel of Fig.~\ref{fig:neg2d} shows the radius of the sphere supporting non-zero negativity, $\tilde{r}_{\Nslash}$, measured in units of the spatial extent of the entangled regions for a 1D and 2D lattice.
It is clear that the entanglement radius grows more slowly on a 2D lattice as $\tilde{r}_{\Nslash}/d_{lat} \sim 0.59(4) d_{lat} $ compared with that in 1D, $\tilde{r}_{\Nslash}/d_{lat} \sim 1.118(1) d_{lat}$.
The further constricted negativity sphere radius expected to be found in a 3D geometry warrants exploration of this non-perturbative lattice effect for reliable calculation of coherent quantum phenomena on both classical and quantum computational frameworks.

A non-vanishing lattice spacing introduces features beyond the finite radius negativity sphere.
The negativity exhibits an oscillatory component with an amplitude that vanishes in the continuum limit,
and falls away from the continuum value as it approaches $\tilde{r}_{\Nslash}$, the surface of the negativity sphere.
These oscillations in negativity introduce an additional systematic error in lattice calculations to be considered,
even within $\tilde{r}_{\Nslash}$.
For small systems, this can lead to orders of magnitude deviations in the negativity from the continuum limit.

The finite value of $\tilde{r}_{\Nslash}$ implies the existence of a non-perturbative reduction in the physical entanglement volume of a lattice calculation if the observable of interest is sensitive to distillable entanglement.
In order to begin understanding the potential
implications of the lattice-spacing-induced finite-sized negativity sphere for LQCD calculations,
we consider relevant lengths scales in a 2D lattice calculation of two \enquote{nucleon-sized} objects interacting through a massless scalar field.

Assuming the nucleon radius is defined by the QCD chiral symmetry breaking scale
$r\chi\sim 1/\Lambda_\chi \sim 0.2~{\rm fm}$,
and the scalar field is defined on a 2D grid with a lattice spacing of $a\sim 0.1~{\rm fm}$
(corresponding to $d_\chi\sim 5$ lattice sites across the nucleon),
the radius of the negativity sphere is $\tilde{r}_{\Nslash}\sim 0.8~{\rm fm}$.
At  this radius, the logarithmic negativity is ${\cal N}\sim 10^{-7}$.
Therefore, beyond a separation of $r_{\Nslash} \sim 1.2~{\rm fm}$, the long-distance entanglement structure of the system is incorrect, but only at the level of $\lsim 10^{-7}$ in the distillable entanglement.
A slightly increased lattice spacing of $a = 0.15~{\rm fm}$
corresponds to a vanishing of the logarithmic negativity at $\tilde{r}_{\Nslash}\sim 0.3~{\rm fm}$,
introducing an entanglement error at the $10^{-5}$ level.
If the size of the nucleons is set by the physical pion mass,
$r\chi\sim 1/m_\pi \sim 1.4~{\rm fm}$ (corresponding to $d_{m_\pi}\sim 30$ lattice sites across for $a = 0.1~{\rm fm}$),
the negativity sphere has a radius of $\tilde{r}_{\Nslash}\sim 47~{\rm fm}$ with ${\cal N}\sim 1\times 10^{-40}$.
In the case of lattice EFT~\cite{Muller:1999cp,Lahde:2019npb,Lee:2020meg,Lee:2008fa} with dynamical pions,
the death of entanglement is likely of greater significance as lattice spacings tend to be larger than those applied in the estimates above.

These 2D estimates indicate that, for coherent quantum observables, LQCD calculations with coarse lattice spacings and quark masses that are physical or heavier may vary in their reliability, with these errors exponentially shrinking with lattice spacing or region pixelation.
Translating the above observations to LQCD and lattice EFT calculations can only be at a qualitative level
without further, {\it in situ}, numerical investigations in 3D.
For LQCD calculations, a much more complex set of estimates are required as the size of the nucleon is dominated by its coupling to pions, which are excitations of the quark condensate,
that behave like a fundamental pseudo-scalar field only at low-energies.

However, many classical observables are likely to be insensitive to a lattice truncation of the negativity, as suggested by smooth two-point functions of the field operators and ($1\times 1$)-site mutual information calculated in a massive scalar field, reflecting continuum structure with only nearest neighbor ($1\times1$)-site negativity~\cite{Audenaert:2002xfl,Botero_2004,kofler2006entanglement,Marcovitch:2008sxc,Calabrese:2009ez,Calabrese:2012ew,MohammadiMozaffar:2017nri,Coser_2017,Klco:2019yrb}.
This perspective aligns well with the successes of
semi-classical
approximations to the structure and interactions of nuclei, including the large-$N_c$ limit of QCD~\cite{tHooft:1973alw,Witten:1980sp,Dashen:1993jt,Dashen:1994qi,Kaplan:1995yg,Kaplan:1996rk}.
Propagating the impact of the non-perturbative negativity sphere to provide a complete quantification of uncertainties for specifically quantum observables requires further research.

Beyond the 2-body negativity spheres, we anticipate irreducible 3-body negativity spheres involving three spatially separated regions that do not factorize into combinations of 1-body and 2-body negativities.
While such irreducibility is well established for qubit systems~\cite{Coffman:1999jd,Dur:2000zz,lohmayer2006entangled,Rangamani:2015qwa},
similar quantities remain to be defined in continuous systems of spatially extended field regions.

In LQCD calculations, Luscher's methods~\cite{Luscher:1986pf,Luscher:1990ux,Luscher:1990ck} have
played a central role in extracting physics from finite-volume, Euclidean-space computations.
These methods are applicable to  simulations on quantum devices with little or no modification, and are expected to play an important role for near-term simulations in small spatial volumes.
Finite lattice spacing artifacts are treated as distinct in LQCD calculations, as they are UV effects, while Luscher's methods are related to the IR structure of the simulation volume.
The impact of the lattice-induced negativity sphere, $\tilde{r}_{\Nslash}$, on finite volume wavefunctions remains to be assessed, though expected to be small for all but quantum coherent observables.

The implications of negativity spheres, $\tilde{r}_{\Nslash}$, in the context of EFTs is interesting to consider further.
At the heart of such effective descriptions are multipole expansions.
This framework enables the effects of extended sources and sinks in QFTs to be included in low-energy EFTs as local operators
($\delta^{(n)}({\bf r})$, $\delta^{\prime (n)}({\bf r})$, ...)
coupled to the dynamical IR degrees of freedom as in, for example, heavy-baryon chiral perturbation theory~\cite{Jenkins:1990jv}.
As EFTs require both an operator structure and a prescribed regularization and renormalization scheme,
extending the effects of the negativity sphere to the EFT suggests that taking renormalization scales that are disparate from the \enquote{size} of the source/sink
could lead to discrepancies in the low-energy description of entanglement.
This point may be relevant to EFT descriptions of baryons~\cite{Jenkins:1990jv} and of
nuclear forces~\cite{Weinberg:1990rz,Weinberg:1991um,PhysRevC.49.2932,Kaplan:1998tg,Epelbaum:2019kcf},
particularly nucleon-nucleon interactions in channels with a tensor force,  which have so far eluded
dimensional regularization~\cite{Beane:2001bc,PhysRevC.74.014003,Kaplan:2019znu}
but can be regulated with smearing in either momentum- or position-space.
In the case of single baryons, it has been suggested that a spatial regularization  would have advantages over massless regularization schemes in the convergence
of chiral expansions of baryon properties~\cite{Young:2005tr}.

In addition to the implications of geometrically-influenced entanglement scales in EFT construction and of negativity spheres in designing latticized calculations of quantum fields,  a perspective remains that detailed understanding of UV and IR entanglement properties can be leveraged for computational advantage~\cite{White:1992zz,vidal2003efficient,schollwock2005density,verstraete2008matrix,Klco:2019yrb}.
While the IR physics of interest may be insensitive to specific choices in the UV structure, demands on computational hardware remain susceptible.
The finite negativity sphere at distance $\tilde{r}_{\Nslash}$ scaling inversely with the pixelation of the field region indicates that UV operators (probing length scales of the lattice spacing) cannot access entangled elements of the field at long distances.
It is possible that this delocalization may be leveraged to improve robustness of quantum hardware against locally interacting, classically correlated sources of noise when simulating QFTs.
Alternatively, designing a field representation with UV entanglement mirroring that found in the IR may provide a reduction in lattice artifacts for coherent quantum observables.

Exploring systematic improvement of the dispersion relation $2 \sin {p_i\over 2} \rightarrow p_i$, reveals that the radius of the lattice negativity sphere, $\tilde{r}_{\Nslash}$, and the geometric decay constant, $\beta$, are essentially unchanged.
Operator and field smearing plays a key role in LQCD calculations,
tempering  UV fluctuations enabling convergence for low-energy quantities,
and mitigating the impact of SO(3) breaking due to
the H(3) spatial lattice.
We have not performed an extensive study of the impact of field or operator smearing, beyond the dispersion relation,
on quantum coherence.

The 2D numerical results provided in this work indicate that the bound on distillable entanglement between two spatially separated regions of the massless non-interacting scalar field vacuum is defined by a decay constant increasing with the dimensionality of spacetime.
Viewed as preliminary evidence of similar properties in more complex gauge theories---such as QCD in which a composite (pseudo)scalar field mediates the long-distance interaction between nucleons---the potential impact of this geometric decay constant in providing an entanglement-sensitive scale in the EFT description of nuclei is discussed.
When pixelating the regions of interest and latticizing the field for non-perturbative calculation, the distillable entanglement is found to suddenly vanish at geometrically large separations (relative to the region size) again dependent on the spatial dimension, becoming more artificially localized in higher dimension.
A thorough and quantitative understanding of the lattice-induced truncation of the distillable entanglement, from the scalar field to QCD, will be a necessary foundation for the reliable calculation of entangled field excitations as well as their propagation to large distances e.g., when probing real-time coherent fragmentation processes, a central target for quantum simulation.
With reduced lattice spacing providing the main source of improvement, the complexity of many-body interactions between collections of lattice sites is determined to be essential for supporting quantum phenomena.
The implications of these geometric features of entanglement in quantum fields, on the convergence of low-energy EFTs and the regulation of spatially extended field objects, sheds new light on objectives to non-perturbatively express non-local quantum effects through a hierarchy of local operators and field elements.

\begin{acknowledgments}
We would like to thank Silas Beane, Ramya Bhaskar, Joe Carlson, David Kaplan, Aidan Murran, Caroline Robin, Kenneth Roche, and Alessandro Roggero for valuable discussions.
We would also like to thank CENPA at the University of Washington for providing an effective work environment over
a period of many months for processing and developing many of the ideas and calculations presented in this paper.
Some of this work was performed on the UW's HYAK High Performance and Data Ecosystem.
We have made extensive use of Wolfram Mathematica~\cite{Mathematica} and the Avanpix multiprecision computing toolbox~\cite{mct2015} for MATLAB~\cite{MATLAB:2020}.
NK and MJS were supported by the Institute for Nuclear Theory with DOE grant No. DE-FG02-00ER41132,
and Fermi National Accelerator Laboratory
PO No. 652197.
This work is supported in part by the U.S. Department of Energy, Office of Science, Office of Advanced Scientific Computing Research (ASCR) quantum algorithm teams program, under field work proposal number ERKJ333.
NK was supported in part by a Microsoft Research PhD Fellowship.
\end{acknowledgments}

\onecolumngrid
\bibliography{halaevbib}

\appendix
\newpage
\section{Figure Tables}
\label{app:datatables}

\begin{table}[h]
\footnotesize
\begin{tabular}{c|c|c}
\hline
\hline
$x$ & $y$ & $\psi_0(x,y)$ \\
\hline
6&0&0.0053146\\
3&1&0.0060582\\
4&1&0.0068117\\
5&1&0.010010\\
6&1&-0.077618\\
7&1&0.0072781\\
8&1&0.0042535\\
9&1&0.0044180\\
2&2&0.0082886\\
3&2&-0.043739\\
4&2&-0.050578\\
5&2&-0.040683\\
6&2&-0.043948\\
7&2&-0.031974\\
8&2&-0.052302\\
9&2&-0.089283\\
10&2&0.013301\\
1&3&0.0057025\\
2&3&-0.034449\\
3&3&-0.046326\\
4&3&-0.027783\\
5&3&0.0019202\\
6&3&0.026263\\
7&3&0.039338\\
8&3&0.022905\\
9&3&-0.028781\\
10&3&-0.089620\\
11&3&0.0012799\\
1&4&0.0067524\\
2&4&-0.041270\\
3&4&-0.030045\\
4&4&0.0076432\\
5&4&0.053102\\
6&4&0.092581\\
7&4&0.11486\\
8&4&0.10603\\
9&4&0.055241\\
\end{tabular}
\begin{tabular}{c|c|c}
10&4&-0.029216\\
11&4&-0.0054380\\
1&5&0.0068186\\
2&5&-0.038103\\
3&5&-0.014644\\
4&5&0.034162\\
5&5&0.089135\\
6&5&0.13717\\
7&5&0.16628\\
8&5&0.16310\\
9&5&0.11490\\
10&5&0.023216\\
11&5&0.0027471\\
0&6&0.0049114\\
1&6&-0.036499\\
2&6&-0.043229\\
3&6&-0.0099229\\
4&6&0.043510\\
5&6&0.10181\\
6&6&0.15268\\
7&6&0.18416\\
8&6&0.18274\\
9&6&0.13402\\
10&6&0.029073\\
11&6&-0.10248\\
12&6&-0.0037411\\
1&7&0.0068186\\
2&7&-0.038103\\
3&7&-0.014644\\
4&7&0.034162\\
5&7&0.089135\\
6&7&0.13717\\
7&7&0.16628\\
8&7&0.16310\\
9&7&0.11490\\
10&7&0.023216\\
11&7&0.0027471\\
1&8&0.0067524\\
\end{tabular}
\begin{tabular}{c|c|c}
2&8&-0.041270\\
3&8&-0.030045\\
4&8&0.0076432\\
5&8&0.053102\\
6&8&0.092581\\
7&8&0.11486\\
8&8&0.10603\\
9&8&0.055241\\
10&8&-0.029216\\
11&8&-0.0054380\\
1&9&0.0057025\\
2&9&-0.034449\\
3&9&-0.046326\\
4&9&-0.027783\\
5&9&0.0019202\\
6&9&0.026263\\
7&9&0.039338\\
8&9&0.022905\\
9&9&-0.028781\\
10&9&-0.089620\\
11&9&0.0012799\\
2&10&0.0082886\\
3&10&-0.043739\\
4&10&-0.050578\\
5&10&-0.040683\\
6&10&-0.043948\\
7&10&-0.031974\\
8&10&-0.052302\\
9&10&-0.089283\\
10&10&0.013301\\
3&11&0.0060582\\
4&11&0.0068117\\
5&11&0.010010\\
6&11&-0.077618\\
7&11&0.0072781\\
8&11&0.0042535\\
9&11&0.0044180\\
6&12&0.0053146\\
\end{tabular}
\begin{tabular}{c|c|c}
32&0&-0.0053146\\
29&1&-0.0044180\\
30&1&-0.0042535\\
31&1&-0.0072781\\
32&1&0.077618\\
33&1&-0.010010\\
34&1&-0.0068117\\
35&1&-0.0060582\\
28&2&-0.013301\\
29&2&0.089283\\
30&2&0.052302\\
31&2&0.031974\\
32&2&0.043948\\
33&2&0.040683\\
34&2&0.050578\\
35&2&0.043739\\
36&2&-0.0082886\\
27&3&-0.0012799\\
28&3&0.089620\\
29&3&0.028781\\
30&3&-0.022905\\
31&3&-0.039338\\
32&3&-0.026263\\
33&3&-0.0019202\\
34&3&0.027783\\
35&3&0.046326\\
36&3&0.034449\\
37&3&-0.0057025\\
27&4&0.0054380\\
28&4&0.029216\\
29&4&-0.055241\\
30&4&-0.10603\\
31&4&-0.11486\\
32&4&-0.092581\\
33&4&-0.053102\\
34&4&-0.0076432\\
35&4&0.030045\\
36&4&0.041270\\
\end{tabular}
\begin{tabular}{c|c|c}
37&4&-0.0067524\\
27&5&-0.0027471\\
28&5&-0.023216\\
29&5&-0.11490\\
30&5&-0.16310\\
31&5&-0.16628\\
32&5&-0.13717\\
33&5&-0.089135\\
34&5&-0.034162\\
35&5&0.014644\\
36&5&0.038103\\
37&5&-0.0068186\\
26&6&0.0037411\\
27&6&0.10248\\
28&6&-0.029073\\
29&6&-0.13402\\
30&6&-0.18274\\
31&6&-0.18416\\
32&6&-0.15268\\
33&6&-0.10181\\
34&6&-0.043510\\
35&6&0.0099229\\
36&6&0.043229\\
37&6&0.036499\\
38&6&-0.0049114\\
27&7&-0.0027471\\
28&7&-0.023216\\
29&7&-0.11490\\
30&7&-0.16310\\
31&7&-0.16628\\
32&7&-0.13717\\
33&7&-0.089135\\
34&7&-0.034162\\
35&7&0.014644\\
36&7&0.038103\\
37&7&-0.0068186\\
27&8&0.0054380\\
28&8&0.029216\\
\end{tabular}
\begin{tabular}{c|c|c}
29&8&-0.055241\\
30&8&-0.10603\\
31&8&-0.11486\\
32&8&-0.092581\\
33&8&-0.053102\\
34&8&-0.0076432\\
35&8&0.030045\\
36&8&0.041270\\
37&8&-0.0067524\\
27&9&-0.0012799\\
28&9&0.089620\\
29&9&0.028781\\
30&9&-0.022905\\
31&9&-0.039338\\
32&9&-0.026263\\
33&9&-0.0019202\\
34&9&0.027783\\
35&9&0.046326\\
36&9&0.034449\\
37&9&-0.0057025\\
28&10&-0.013301\\
29&10&0.089283\\
30&10&0.052302\\
31&10&0.031974\\
32&10&0.043948\\
33&10&0.040683\\
34&10&0.050578\\
35&10&0.043739\\
36&10&-0.0082886\\
29&11&-0.0044180\\
30&11&-0.0042535\\
31&11&-0.0072781\\
32&11&0.077618\\
33&11&-0.010010\\
34&11&-0.0068117\\
35&11&-0.0060582\\
32&12&-0.0053146\\
\hline
\hline
\end{tabular}
  \caption{Wavefunction amplitudes shown in the left panel of Fig.~\ref{fig:GHgammaGroundState13}.}
  \label{tab:GHgammaGroundState13sep13}
\end{table}

\begin{table}
\begin{tabular}{c|c|c}
\hline
\hline
$x$ & $y$ & $\psi_0(x,y)$ \\
\hline
6&0& $4.9034\times 10^{-8}$ \\
3&1& $-5.2388\times 10^{-9}$ \\
4&1& $-8.1054\times 10^{-8}$ \\
5&1& $-1.3491\times 10^{-6}$ \\
6&1& $6.5912\times 10^{-6}$ \\
7&1& $-1.3237\times 10^{-6}$ \\
8&1& $-6.7897\times 10^{-8}$ \\
9&1& $-3.8544\times 10^{-9}$ \\
2&2& $-8.6614\times 10^{-9}$ \\
3&2& $-2.7111\times 10^{-7}$ \\
4&2& $-8.1358\times 10^{-6}$ \\
5&2& $0.00019040$ \\
6&2& $-0.00051386$ \\
7&2& $0.00018701$ \\
8&2& $-8.9529\times 10^{-6}$ \\
9&2& $-2.5605\times 10^{-7}$ \\
10&2& $-5.1844\times 10^{-9}$ \\
1&3& $-5.3318\times 10^{-9}$ \\
2&3& $-2.7117\times 10^{-7}$ \\
3&3& $-0.000013289$ \\
4&3& $0.00061389$ \\
5&3& $-0.0045493$ \\
6&3& $0.0092643$ \\
7&3& $-0.0044675$ \\
8&3& $0.00061098$ \\
9&3& $-0.000015291$ \\
10&3& $-2.5654\times 10^{-7}$ \\
\end{tabular}
\begin{tabular}{c|c|c}
11&3& $-3.1143\times 10^{-9}$ \\
1&4& $-8.2163\times 10^{-8}$ \\
2&4& $-8.0923\times 10^{-6}$ \\
3&4& $0.00061368$ \\
4&4& $-0.0083757$ \\
5&4& $0.039844$ \\
6&4& $-0.069559$ \\
7&4& $0.039077$ \\
8&4& $-0.0081646$ \\
9&4& $0.00060715$ \\
10&4& $-9.2898\times 10^{-6}$ \\
11&4& $-5.7750\times 10^{-8}$ \\
1&5& $-1.3578\times 10^{-6}$ \\
2&5& $0.00019088$ \\
3&5& $-0.0045545$ \\
4&5& $0.039877$ \\
5&5& $-0.15121$ \\
6&5& $0.24273$ \\
7&5& $-0.14799$ \\
8&5& $0.038412$ \\
9&5& $-0.0043252$ \\
10&5& $0.00017830$ \\
11&5& $-1.2394\times 10^{-6}$ \\
0&6& $4.4273\times 10^{-8}$ \\
1&6& $7.0459\times 10^{-6}$ \\
2&6& $-0.00052161$ \\
3&6& $0.0093308$ \\
4&6& $-0.069963$ \\
\end{tabular}
\begin{tabular}{c|c|c}
5&6& $0.24398$ \\
6&6& $-0.37969$ \\
7&6& $0.23827$ \\
8&6& $-0.066852$ \\
9&6& $0.0086482$ \\
10&6& $-0.00045085$ \\
11&6& $4.0623\times 10^{-6}$ \\
12&6& $6.6846\times 10^{-8}$ \\
1&7& $-1.3578\times 10^{-6}$ \\
2&7& $0.00019088$ \\
3&7& $-0.0045545$ \\
4&7& $0.039877$ \\
5&7& $-0.15121$ \\
6&7& $0.24273$ \\
7&7& $-0.14799$ \\
8&7& $0.038412$ \\
9&7& $-0.0043252$ \\
10&7& $0.00017830$ \\
11&7& $-1.2394\times 10^{-6}$ \\
1&8& $-8.2163\times 10^{-8}$ \\
2&8& $-8.0923\times 10^{-6}$ \\
3&8& $0.00061368$ \\
4&8& $-0.0083757$ \\
5&8& $0.039844$ \\
6&8& $-0.069559$ \\
7&8& $0.039077$ \\
8&8& $-0.0081646$ \\
9&8& $0.00060715$ \\
10&8& $-9.2898\times 10^{-6}$ \\
\end{tabular}
\begin{tabular}{c|c|c}
11&8& $-5.7750\times 10^{-8}$ \\
1&9& $-5.3318\times 10^{-9}$ \\
2&9& $-2.7117\times 10^{-7}$ \\
3&9& $-0.000013289$ \\
4&9& $0.00061389$ \\
5&9& $-0.0045493$ \\
6&9& $0.0092643$ \\
7&9& $-0.0044675$ \\
8&9& $0.00061098$ \\
9&9& $-0.000015291$ \\
10&9& $-2.5654\times 10^{-7}$ \\
11&9& $-3.1143\times 10^{-9}$ \\
2&10& $-8.6614\times 10^{-9}$ \\
3&10& $-2.7111\times 10^{-7}$ \\
4&10& $-8.1358\times 10^{-6}$ \\
5&10& $0.00019040$ \\
6&10& $-0.00051386$ \\
7&10& $0.00018701$ \\
8&10& $-8.9529\times 10^{-6}$ \\
9&10& $-2.5605\times 10^{-7}$ \\
10&10& $-5.1844\times 10^{-9}$ \\
3&11& $-5.2388\times 10^{-9}$ \\
4&11& $-8.1054\times 10^{-8}$ \\
5&11& $-1.3491\times 10^{-6}$ \\
6&11& $6.5912\times 10^{-6}$ \\
7&11& $-1.3237\times 10^{-6}$ \\
8&11& $-6.7897\times 10^{-8}$ \\
9&11& $-3.8544\times 10^{-9}$ \\
6&12& $4.9034\times 10^{-8}$ \\
\hline
\hline
\end{tabular}
  \caption{Wavefunction amplitudes shown in the right panel of Fig.~\ref{fig:GHgammaGroundState13}.}
  \label{tab:GHgammaGroundState13sep100}
\end{table}

\begin{table}
\begin{tabular}{c|c}
\multicolumn{2}{c}{$d_{lat} = 7$} \\
\hline
\hline
$\tilde{r}/d_{avg}$& $\mathcal{N}$ \\
\hline
\hline
0.553 & $ 1.20\times 10^{-3} $ \\
0.737 & $ 5.90\times 10^{-4} $ \\
0.922 & $ 2.58\times 10^{-4} $ \\
1.11 & $ 9.09\times 10^{-5} $ \\
1.29 & $ 2.29\times 10^{-5} $ \\
1.47 & $ 6.81\times 10^{-6} $ \\
1.66 & $ 2.91\times 10^{-6} $ \\
1.84 & $ 1.32\times 10^{-6} $ \\
2.03 & $ 5.12\times 10^{-7} $ \\
2.21 & $ 1.57\times 10^{-7} $ \\
2.4 & $ 4.84\times 10^{-8} $ \\
2.58 & $ 1.29\times 10^{-8} $ \\
2.77 & $ 2.41\times 10^{-9} $ \\
2.95 & $ 7.82\times 10^{-10} $ \\
3.13 & $ 3.66\times 10^{-10} $ \\
3.32 & $ 1.83\times 10^{-10} $ \\
3.5 & $ 8.01\times 10^{-11} $ \\
3.69 & $ 1.32\times 10^{-11} $ \\
\hline
\hline
\end{tabular}
\begin{tabular}{c|c}
\multicolumn{2}{c}{$d_{lat} = 9$} \\
\hline
\hline
$\tilde{r}/d_{avg}$& $\mathcal{N}$ \\
\hline
\hline
0.553 & $ 1.47\times 10^{-3} $ \\
0.691 & $ 7.57\times 10^{-4} $ \\
0.829 & $ 3.46\times 10^{-4} $ \\
0.968 & $ 1.34\times 10^{-4} $ \\
1.11 & $ 5.16\times 10^{-5} $ \\
1.24 & $ 2.47\times 10^{-5} $ \\
1.38 & $ 1.38\times 10^{-5} $ \\
1.52 & $ 8.13\times 10^{-6} $ \\
1.66 & $ 4.78\times 10^{-6} $ \\
1.8 & $ 2.67\times 10^{-6} $ \\
1.94 & $ 1.36\times 10^{-6} $ \\
2.07 & $ 6.15\times 10^{-7} $ \\
2.21 & $ 2.25\times 10^{-7} $ \\
2.35 & $ 6.57\times 10^{-8} $ \\
2.49 & $ 2.40\times 10^{-8} $ \\
2.63 & $ 1.18\times 10^{-8} $ \\
2.76 & $ 6.68\times 10^{-9} $ \\
2.9 & $ 4.01\times 10^{-9} $ \\
3.04 & $ 2.39\times 10^{-9} $ \\
3.18 & $ 1.33\times 10^{-9} $ \\
3.32 & $ 6.03\times 10^{-10} $ \\
3.46 & $ 1.80\times 10^{-10} $ \\
3.59 & $ 5.06\times 10^{-11} $ \\
3.73 & $ 1.69\times 10^{-11} $ \\
3.87 & $ 4.61\times 10^{-12} $ \\
4.01 & $ 1.13\times 10^{-12} $ \\
4.15 & $ 4.72\times 10^{-13} $ \\
4.29 & $ 2.56\times 10^{-13} $ \\
4.42 & $ 1.52\times 10^{-13} $ \\
4.56 & $ 9.09\times 10^{-14} $ \\
4.7 & $ 5.05\times 10^{-14} $ \\
4.84 & $ 2.17\times 10^{-14} $ \\
4.98 & $ 1.39\times 10^{-16} $ \\
\hline
\hline
\end{tabular}
\begin{tabular}{c|c}
\multicolumn{2}{c}{$d_{lat} = 11$} \\
\hline
\hline
$\tilde{r}/d_{avg}$& $\mathcal{N}$ \\
\hline
\hline
0.106 & $ 2.35\times 10^{-2} $ \\
0.211 & $ 9.16\times 10^{-3} $ \\
0.317 & $ 5.47\times 10^{-3} $ \\
0.423 & $ 3.40\times 10^{-3} $ \\
0.528 & $ 2.11\times 10^{-3} $ \\
0.634 & $ 1.27\times 10^{-3} $ \\
0.74 & $ 7.16\times 10^{-4} $ \\
0.845 & $ 3.69\times 10^{-4} $ \\
0.951 & $ 1.78\times 10^{-4} $ \\
1.06 & $ 8.98\times 10^{-5} $ \\
1.16 & $ 5.09\times 10^{-5} $ \\
1.27 & $ 3.14\times 10^{-5} $ \\
1.37 & $ 2.03\times 10^{-5} $ \\
1.48 & $ 1.32\times 10^{-5} $ \\
1.58 & $ 8.46\times 10^{-6} $ \\
1.69 & $ 5.21\times 10^{-6} $ \\
1.8 & $ 3.02\times 10^{-6} $ \\
1.9 & $ 1.61\times 10^{-6} $ \\
2.01 & $ 7.76\times 10^{-7} $ \\
2.11 & $ 3.41\times 10^{-7} $ \\
2.22 & $ 1.48\times 10^{-7} $ \\
2.32 & $ 7.27\times 10^{-8} $ \\
2.43 & $ 4.12\times 10^{-8} $ \\
2.54 & $ 2.57\times 10^{-8} $ \\
2.64 & $ 1.68\times 10^{-8} $ \\
2.75 & $ 1.12\times 10^{-8} $ \\
2.85 & $ 7.43\times 10^{-9} $ \\
2.96 & $ 4.77\times 10^{-9} $ \\
\end{tabular}
\begin{tabular}{c|c}
3.06 & $ 2.91\times 10^{-9} $ \\
3.17 & $ 1.68\times 10^{-9} $ \\
3.28 & $ 9.19\times 10^{-10} $ \\
3.38 & $ 4.81\times 10^{-10} $ \\
3.49 & $ 2.35\times 10^{-10} $ \\
3.59 & $ 1.06\times 10^{-10} $ \\
3.7 & $ 4.99\times 10^{-11} $ \\
3.8 & $ 2.73\times 10^{-11} $ \\
3.91 & $ 1.68\times 10^{-11} $ \\
4.01 & $ 1.11\times 10^{-11} $ \\
4.12 & $ 7.60\times 10^{-12} $ \\
4.23 & $ 5.23\times 10^{-12} $ \\
4.33 & $ 3.53\times 10^{-12} $ \\
4.44 & $ 2.28\times 10^{-12} $ \\
4.54 & $ 1.34\times 10^{-12} $ \\
4.65 & $ 6.58\times 10^{-13} $ \\
4.75 & $ 2.27\times 10^{-13} $ \\
4.86 & $ 4.55\times 10^{-14} $ \\
4.97 & $ 1.21\times 10^{-14} $ \\
5.07 & $ 4.39\times 10^{-15} $ \\
5.18 & $ 1.43\times 10^{-15} $ \\
5.28 & $ 3.71\times 10^{-16} $ \\
5.39 & $ 1.45\times 10^{-16} $ \\
5.49 & $ 7.89\times 10^{-17} $ \\
5.6 & $ 4.87\times 10^{-17} $ \\
5.71 & $ 3.17\times 10^{-17} $ \\
5.81 & $ 2.08\times 10^{-17} $ \\
5.92 & $ 1.32\times 10^{-17} $ \\
6.02 & $ 7.58\times 10^{-18} $ \\
6.13 & $ 3.27\times 10^{-18} $ \\
\hline
\hline
\end{tabular}
  \caption{Logarithmic negativity as a function of separation measured in units of the averaged region diameter with an N-type boundary appearing in the left panel of Fig.~\ref{fig:neg2d}.}
  \label{tab:2dnegativity1}
\end{table}

\begin{table}
\centering
\footnotesize
\begin{tabular}{c|c}
\multicolumn{2}{c}{$d_{lat} = 13$} \\
\hline
\hline
$\tilde{r}/d_{avg}$& $\mathcal{N}$ \\
\hline
\hline
0.0885 & $ 2.90\times 10^{-2} $ \\
0.177 & $ 1.27\times 10^{-2} $ \\
0.266 & $ 7.93\times 10^{-3} $ \\
0.354 & $ 5.17\times 10^{-3} $ \\
0.443 & $ 3.40\times 10^{-3} $ \\
0.531 & $ 2.20\times 10^{-3} $ \\
0.62 & $ 1.37\times 10^{-3} $ \\
0.708 & $ 8.06\times 10^{-4} $ \\
0.797 & $ 4.46\times 10^{-4} $ \\
0.885 & $ 2.42\times 10^{-4} $ \\
0.974 & $ 1.38\times 10^{-4} $ \\
1.06 & $ 8.51\times 10^{-5} $ \\
1.15 & $ 5.54\times 10^{-5} $ \\
1.24 & $ 3.73\times 10^{-5} $ \\
1.33 & $ 2.53\times 10^{-5} $ \\
1.42 & $ 1.71\times 10^{-5} $ \\
1.5 & $ 1.13\times 10^{-5} $ \\
1.59 & $ 7.27\times 10^{-6} $ \\
1.68 & $ 4.46\times 10^{-6} $ \\
1.77 & $ 2.62\times 10^{-6} $ \\
1.86 & $ 1.49\times 10^{-6} $ \\
1.95 & $ 8.45\times 10^{-7} $ \\
2.04 & $ 4.90\times 10^{-7} $ \\
2.12 & $ 2.96\times 10^{-7} $ \\
2.21 & $ 1.88\times 10^{-7} $ \\
2.3 & $ 1.24\times 10^{-7} $ \\
2.39 & $ 8.46\times 10^{-8} $ \\
2.48 & $ 5.85\times 10^{-8} $ \\
2.57 & $ 4.06\times 10^{-8} $ \\
2.66 & $ 2.79\times 10^{-8} $ \\
2.74 & $ 1.88\times 10^{-8} $ \\
2.83 & $ 1.23\times 10^{-8} $ \\
2.92 & $ 7.80\times 10^{-9} $ \\
3.01 & $ 4.72\times 10^{-9} $ \\
3.1 & $ 2.73\times 10^{-9} $ \\
3.19 & $ 1.52\times 10^{-9} $ \\
3.28 & $ 8.48\times 10^{-10} $ \\
3.36 & $ 4.96\times 10^{-10} $ \\
3.45 & $ 3.07\times 10^{-10} $ \\
3.54 & $ 1.97\times 10^{-10} $ \\
3.63 & $ 1.29\times 10^{-10} $ \\
\end{tabular}
\begin{tabular}{c|c}
3.72 & $ 8.43\times 10^{-11} $ \\
3.81 & $ 5.45\times 10^{-11} $ \\
3.89 & $ 3.45\times 10^{-11} $ \\
3.98 & $ 2.15\times 10^{-11} $ \\
4.07 & $ 1.32\times 10^{-11} $ \\
4.16 & $ 7.89\times 10^{-12} $ \\
4.25 & $ 4.55\times 10^{-12} $ \\
4.34 & $ 2.52\times 10^{-12} $ \\
4.43 & $ 1.36\times 10^{-12} $ \\
4.51 & $ 7.28\times 10^{-13} $ \\
4.6 & $ 3.76\times 10^{-13} $ \\
4.69 & $ 1.85\times 10^{-13} $ \\
4.78 & $ 9.15\times 10^{-14} $ \\
4.87 & $ 5.04\times 10^{-14} $ \\
4.96 & $ 3.11\times 10^{-14} $ \\
5.05 & $ 2.08\times 10^{-14} $ \\
5.13 & $ 1.45\times 10^{-14} $ \\
5.22 & $ 1.04\times 10^{-14} $ \\
5.31 & $ 7.45\times 10^{-15} $ \\
5.4 & $ 5.30\times 10^{-15} $ \\
5.49 & $ 3.67\times 10^{-15} $ \\
5.58 & $ 2.41\times 10^{-15} $ \\
5.67 & $ 1.46\times 10^{-15} $ \\
5.75 & $ 7.89\times 10^{-16} $ \\
5.84 & $ 3.85\times 10^{-16} $ \\
5.93 & $ 1.71\times 10^{-16} $ \\
6.02 & $ 6.10\times 10^{-17} $ \\
6.11 & $ 1.64\times 10^{-17} $ \\
6.2 & $ 6.17\times 10^{-18} $ \\
6.28 & $ 3.19\times 10^{-18} $ \\
6.37 & $ 1.87\times 10^{-18} $ \\
6.46 & $ 1.13\times 10^{-18} $ \\
6.55 & $ 6.74\times 10^{-19} $ \\
6.64 & $ 3.68\times 10^{-19} $ \\
6.73 & $ 1.70\times 10^{-19} $ \\
6.82 & $ 7.10\times 10^{-20} $ \\
6.9 & $ 3.36\times 10^{-20} $ \\
6.99 & $ 1.81\times 10^{-20} $ \\
7.08 & $ 1.02\times 10^{-20} $ \\
7.17 & $ 5.45\times 10^{-21} $ \\
7.26 & $ 2.35\times 10^{-21} $ \\
7.35 & $ 1.61\times 10^{-22} $ \\
\hline
\hline
\end{tabular}
\begin{tabular}{c|c}
\multicolumn{2}{c}{$d_{lat} = 15$} \\
\hline
\hline
$\tilde{r}/d_{avg}$& $\mathcal{N}$ \\
\hline
\hline
0.0765 & $ 3.32\times 10^{-2} $ \\
0.153 & $ 1.55\times 10^{-2} $ \\
0.23 & $ 1.00\times 10^{-2} $ \\
0.306 & $ 6.72\times 10^{-3} $ \\
0.383 & $ 4.57\times 10^{-3} $ \\
0.459 & $ 3.08\times 10^{-3} $ \\
0.536 & $ 2.03\times 10^{-3} $ \\
0.612 & $ 1.28\times 10^{-3} $ \\
0.689 & $ 7.78\times 10^{-4} $ \\
0.765 & $ 4.60\times 10^{-4} $ \\
0.842 & $ 2.77\times 10^{-4} $ \\
0.918 & $ 1.75\times 10^{-4} $ \\
0.995 & $ 1.17\times 10^{-4} $ \\
1.07 & $ 8.05\times 10^{-5} $ \\
1.15 & $ 5.67\times 10^{-5} $ \\
1.22 & $ 4.02\times 10^{-5} $ \\
1.3 & $ 2.84\times 10^{-5} $ \\
1.38 & $ 1.98\times 10^{-5} $ \\
1.45 & $ 1.35\times 10^{-5} $ \\
1.53 & $ 8.94\times 10^{-6} $ \\
1.61 & $ 5.73\times 10^{-6} $ \\
1.68 & $ 3.56\times 10^{-6} $ \\
1.76 & $ 2.18\times 10^{-6} $ \\
1.84 & $ 1.36\times 10^{-6} $ \\
1.91 & $ 8.75\times 10^{-7} $ \\
1.99 & $ 5.89\times 10^{-7} $ \\
2.07 & $ 4.10\times 10^{-7} $ \\
2.14 & $ 2.92\times 10^{-7} $ \\
2.22 & $ 2.10\times 10^{-7} $ \\
2.3 & $ 1.52\times 10^{-7} $ \\
2.37 & $ 1.10\times 10^{-7} $ \\
2.45 & $ 7.81\times 10^{-8} $ \\
2.52 & $ 5.46\times 10^{-8} $ \\
2.6 & $ 3.74\times 10^{-8} $ \\
2.68 & $ 2.49\times 10^{-8} $ \\
2.75 & $ 1.61\times 10^{-8} $ \\
2.83 & $ 1.01\times 10^{-8} $ \\
\end{tabular}
\begin{tabular}{c|c}
2.91 & $ 6.19\times 10^{-9} $ \\
2.98 & $ 3.75\times 10^{-9} $ \\
3.06 & $ 2.32\times 10^{-9} $ \\
3.14 & $ 1.50\times 10^{-9} $ \\
3.21 & $ 1.01\times 10^{-9} $ \\
3.29 & $ 6.99\times 10^{-10} $ \\
3.37 & $ 4.91\times 10^{-10} $ \\
3.44 & $ 3.47\times 10^{-10} $ \\
3.52 & $ 2.43\times 10^{-10} $ \\
3.6 & $ 1.68\times 10^{-10} $ \\
3.67 & $ 1.14\times 10^{-10} $ \\
3.75 & $ 7.67\times 10^{-11} $ \\
3.83 & $ 5.10\times 10^{-11} $ \\
3.9 & $ 3.37\times 10^{-11} $ \\
3.98 & $ 2.21\times 10^{-11} $ \\
4.05 & $ 1.43\times 10^{-11} $ \\
4.13 & $ 9.09\times 10^{-12} $ \\
4.21 & $ 5.65\times 10^{-12} $ \\
4.28 & $ 3.47\times 10^{-12} $ \\
4.36 & $ 2.15\times 10^{-12} $ \\
4.44 & $ 1.39\times 10^{-12} $ \\
4.51 & $ 9.32\times 10^{-13} $ \\
4.59 & $ 6.51\times 10^{-13} $ \\
4.67 & $ 4.67\times 10^{-13} $ \\
4.74 & $ 3.40\times 10^{-13} $ \\
4.82 & $ 2.49\times 10^{-13} $ \\
4.9 & $ 1.82\times 10^{-13} $ \\
4.97 & $ 1.31\times 10^{-13} $ \\
5.05 & $ 9.22\times 10^{-14} $ \\
5.13 & $ 6.28\times 10^{-14} $ \\
5.2 & $ 4.13\times 10^{-14} $ \\
5.28 & $ 2.62\times 10^{-14} $ \\
5.36 & $ 1.62\times 10^{-14} $ \\
5.43 & $ 9.91\times 10^{-15} $ \\
5.51 & $ 5.94\times 10^{-15} $ \\
5.59 & $ 3.44\times 10^{-15} $ \\
5.66 & $ 1.90\times 10^{-15} $ \\
\end{tabular}
\begin{tabular}{c|c}
5.74 & $ 1.00\times 10^{-15} $ \\
5.81 & $ 5.14\times 10^{-16} $ \\
5.89 & $ 2.58\times 10^{-16} $ \\
5.97 & $ 1.30\times 10^{-16} $ \\
6.04 & $ 6.94\times 10^{-17} $ \\
6.12 & $ 4.10\times 10^{-17} $ \\
6.2 & $ 2.65\times 10^{-17} $ \\
6.27 & $ 1.82\times 10^{-17} $ \\
6.35 & $ 1.30\times 10^{-17} $ \\
6.43 & $ 9.51\times 10^{-18} $ \\
6.5 & $ 6.99\times 10^{-18} $ \\
6.58 & $ 5.12\times 10^{-18} $ \\
6.66 & $ 3.68\times 10^{-18} $ \\
6.73 & $ 2.55\times 10^{-18} $ \\
6.81 & $ 1.69\times 10^{-18} $ \\
6.89 & $ 1.04\times 10^{-18} $ \\
6.96 & $ 5.96\times 10^{-19} $ \\
7.04 & $ 3.28\times 10^{-19} $ \\
7.12 & $ 1.75\times 10^{-19} $ \\
7.19 & $ 8.57\times 10^{-20} $ \\
7.27 & $ 3.49\times 10^{-20} $ \\
7.34 & $ 1.28\times 10^{-20} $ \\
7.42 & $ 6.08\times 10^{-21} $ \\
7.5 & $ 3.60\times 10^{-21} $ \\
7.57 & $ 2.39\times 10^{-21} $ \\
7.65 & $ 1.68\times 10^{-21} $ \\
7.73 & $ 1.22\times 10^{-21} $ \\
7.8 & $ 8.96\times 10^{-22} $ \\
7.88 & $ 6.57\times 10^{-22} $ \\
7.96 & $ 4.74\times 10^{-22} $ \\
8.03 & $ 3.29\times 10^{-22} $ \\
8.11 & $ 2.13\times 10^{-22} $ \\
8.19 & $ 1.18\times 10^{-22} $ \\
8.26 & $ 4.37\times 10^{-23} $ \\
8.34 & $ 8.04\times 10^{-24} $ \\
8.42 & $ 2.08\times 10^{-24} $ \\
8.49 & $ 4.51\times 10^{-25} $ \\
\hline
\hline
\end{tabular}
  \caption{Logarithmic negativity as a function of separation measured in units of the averaged region diameter with an N-type boundary appearing in the left panel of Fig.~\ref{fig:neg2d}.}
  \label{tab:2dnegativity2}
\end{table}

\begin{table}
\footnotesize
\begin{tabular}{c|c}
\multicolumn{2}{c}{$d_{lat} = 17$} \\
\hline
\hline
$\tilde{r}/d_{avg}$& $\mathcal{N}$ \\
\hline
\hline
0.0661 & $ 3.98\times 10^{-2} $ \\
0.132 & $ 2.01\times 10^{-2} $ \\
0.198 & $ 1.34\times 10^{-2} $ \\
0.264 & $ 9.33\times 10^{-3} $ \\
0.33 & $ 6.60\times 10^{-3} $ \\
0.397 & $ 4.68\times 10^{-3} $ \\
0.463 & $ 3.29\times 10^{-3} $ \\
0.529 & $ 2.26\times 10^{-3} $ \\
0.595 & $ 1.50\times 10^{-3} $ \\
0.661 & $ 9.66\times 10^{-4} $ \\
0.727 & $ 6.07\times 10^{-4} $ \\
0.793 & $ 3.85\times 10^{-4} $ \\
0.859 & $ 2.53\times 10^{-4} $ \\
0.925 & $ 1.74\times 10^{-4} $ \\
0.991 & $ 1.24\times 10^{-4} $ \\
1.06 & $ 9.10\times 10^{-5} $ \\
1.12 & $ 6.73\times 10^{-5} $ \\
1.19 & $ 5.00\times 10^{-5} $ \\
1.26 & $ 3.70\times 10^{-5} $ \\
1.32 & $ 2.70\times 10^{-5} $ \\
1.39 & $ 1.94\times 10^{-5} $ \\
1.45 & $ 1.37\times 10^{-5} $ \\
1.52 & $ 9.36\times 10^{-6} $ \\
1.59 & $ 6.24\times 10^{-6} $ \\
1.65 & $ 4.09\times 10^{-6} $ \\
1.72 & $ 2.69\times 10^{-6} $ \\
1.78 & $ 1.81\times 10^{-6} $ \\
1.85 & $ 1.26\times 10^{-6} $ \\
1.92 & $ 9.04\times 10^{-7} $ \\
1.98 & $ 6.66\times 10^{-7} $ \\
2.05 & $ 4.98\times 10^{-7} $ \\
2.11 & $ 3.76\times 10^{-7} $ \\
2.18 & $ 2.85\times 10^{-7} $ \\
2.25 & $ 2.15\times 10^{-7} $ \\
2.31 & $ 1.60\times 10^{-7} $ \\
2.38 & $ 1.18\times 10^{-7} $ \\
\end{tabular}
\begin{tabular}{c|c}
2.45 & $ 8.51\times 10^{-8} $ \\
2.51 & $ 6.00\times 10^{-8} $ \\
2.58 & $ 4.13\times 10^{-8} $ \\
2.64 & $ 2.77\times 10^{-8} $ \\
2.71 & $ 1.80\times 10^{-8} $ \\
2.78 & $ 1.14\times 10^{-8} $ \\
2.84 & $ 7.09\times 10^{-9} $ \\
2.91 & $ 4.46\times 10^{-9} $ \\
2.97 & $ 2.92\times 10^{-9} $ \\
3.04 & $ 2.00\times 10^{-9} $ \\
3.11 & $ 1.43\times 10^{-9} $ \\
3.17 & $ 1.06\times 10^{-9} $ \\
3.24 & $ 7.97\times 10^{-10} $ \\
3.3 & $ 6.07\times 10^{-10} $ \\
3.37 & $ 4.65\times 10^{-10} $ \\
3.44 & $ 3.55\times 10^{-10} $ \\
3.5 & $ 2.69\times 10^{-10} $ \\
3.57 & $ 2.01\times 10^{-10} $ \\
3.63 & $ 1.47\times 10^{-10} $ \\
3.7 & $ 1.05\times 10^{-10} $ \\
3.77 & $ 7.28\times 10^{-11} $ \\
3.83 & $ 4.92\times 10^{-11} $ \\
3.9 & $ 3.25\times 10^{-11} $ \\
3.97 & $ 2.10\times 10^{-11} $ \\
4.03 & $ 1.35\times 10^{-11} $ \\
4.1 & $ 8.66\times 10^{-12} $ \\
4.16 & $ 5.61\times 10^{-12} $ \\
4.23 & $ 3.70\times 10^{-12} $ \\
4.3 & $ 2.50\times 10^{-12} $ \\
4.36 & $ 1.73\times 10^{-12} $ \\
4.43 & $ 1.24\times 10^{-12} $ \\
4.49 & $ 9.05\times 10^{-13} $ \\
4.56 & $ 6.75\times 10^{-13} $ \\
4.63 & $ 5.10\times 10^{-13} $ \\
4.69 & $ 3.89\times 10^{-13} $ \\
4.76 & $ 2.96\times 10^{-13} $ \\
4.82 & $ 2.24\times 10^{-13} $ \\
\end{tabular}
\begin{tabular}{c|c}
4.89 & $ 1.68\times 10^{-13} $ \\
4.96 & $ 1.24\times 10^{-13} $ \\
5.02 & $ 8.98\times 10^{-14} $ \\
5.09 & $ 6.33\times 10^{-14} $ \\
5.16 & $ 4.37\times 10^{-14} $ \\
5.22 & $ 2.94\times 10^{-14} $ \\
5.29 & $ 1.94\times 10^{-14} $ \\
5.35 & $ 1.25\times 10^{-14} $ \\
5.42 & $ 7.81\times 10^{-15} $ \\
5.49 & $ 4.77\times 10^{-15} $ \\
5.55 & $ 2.92\times 10^{-15} $ \\
5.62 & $ 1.86\times 10^{-15} $ \\
5.68 & $ 1.24\times 10^{-15} $ \\
5.75 & $ 8.70\times 10^{-16} $ \\
5.82 & $ 6.31\times 10^{-16} $ \\
5.88 & $ 4.68\times 10^{-16} $ \\
5.95 & $ 3.51\times 10^{-16} $ \\
6.01 & $ 2.65\times 10^{-16} $ \\
6.08 & $ 1.99\times 10^{-16} $ \\
6.15 & $ 1.47\times 10^{-16} $ \\
6.21 & $ 1.07\times 10^{-16} $ \\
6.28 & $ 7.54\times 10^{-17} $ \\
6.34 & $ 5.12\times 10^{-17} $ \\
6.41 & $ 3.35\times 10^{-17} $ \\
6.48 & $ 2.13\times 10^{-17} $ \\
6.54 & $ 1.34\times 10^{-17} $ \\
6.61 & $ 8.39\times 10^{-18} $ \\
6.68 & $ 5.28\times 10^{-18} $ \\
6.74 & $ 3.32\times 10^{-18} $ \\
6.81 & $ 2.08\times 10^{-18} $ \\
6.87 & $ 1.30\times 10^{-18} $ \\
6.94 & $ 8.19\times 10^{-19} $ \\
7.01 & $ 5.17\times 10^{-19} $ \\
7.07 & $ 3.26\times 10^{-19} $ \\
7.14 & $ 2.04\times 10^{-19} $ \\
7.2 & $ 1.25\times 10^{-19} $ \\
7.27 & $ 7.43\times 10^{-20} $ \\
\end{tabular}
\begin{tabular}{c|c}
7.34 & $ 4.37\times 10^{-20} $ \\
7.4 & $ 2.60\times 10^{-20} $ \\
7.47 & $ 1.61\times 10^{-20} $ \\
7.53 & $ 1.04\times 10^{-20} $ \\
7.6 & $ 7.05\times 10^{-21} $ \\
7.67 & $ 4.88\times 10^{-21} $ \\
7.73 & $ 3.41\times 10^{-21} $ \\
7.8 & $ 2.39\times 10^{-21} $ \\
7.86 & $ 1.65\times 10^{-21} $ \\
7.93 & $ 1.12\times 10^{-21} $ \\
8. & $ 7.44\times 10^{-22} $ \\
8.06 & $ 4.79\times 10^{-22} $ \\
8.13 & $ 3.00\times 10^{-22} $ \\
8.2 & $ 1.84\times 10^{-22} $ \\
8.26 & $ 1.08\times 10^{-22} $ \\
8.33 & $ 5.97\times 10^{-23} $ \\
8.39 & $ 2.91\times 10^{-23} $ \\
8.46 & $ 1.25\times 10^{-23} $ \\
8.53 & $ 5.85\times 10^{-24} $ \\
8.59 & $ 3.31\times 10^{-24} $ \\
8.66 & $ 2.14\times 10^{-24} $ \\
8.72 & $ 1.49\times 10^{-24} $ \\
8.79 & $ 1.08\times 10^{-24} $ \\
8.86 & $ 7.99\times 10^{-25} $ \\
8.92 & $ 5.96\times 10^{-25} $ \\
8.99 & $ 4.44\times 10^{-25} $ \\
9.05 & $ 3.25\times 10^{-25} $ \\
9.12 & $ 2.30\times 10^{-25} $ \\
9.19 & $ 1.54\times 10^{-25} $ \\
9.25 & $ 9.20\times 10^{-26} $ \\
9.32 & $ 4.39\times 10^{-26} $ \\
9.38 & $ 1.04\times 10^{-26} $ \\
9.45 & $ 1.06\times 10^{-27} $ \\
9.52 & $ 3.19\times 10^{-28} $ \\
9.58 & $ 1.11\times 10^{-28} $ \\
9.65 & $ 1.39\times 10^{-29} $ \\
\hline
\hline
\end{tabular}
  \caption{Logarithmic negativity as a function of separation measured in units of the averaged region diameter with an N-type boundary appearing in the left panel of Fig.~\ref{fig:neg2d}.}
  \label{tab:2dnegativity3}
\end{table}

\begin{table}
\begin{tabular}{c|c}
\multicolumn{2}{c}{$d_{lat} = 19$} \\
\hline
\hline
$\tilde{r}/d_{avg}$& $\mathcal{N}$ \\
\hline
\hline
0.058 & $ 4.98\times 10^{-2} $ \\
0.116 & $ 2.61\times 10^{-2} $ \\
0.174 & $ 1.75\times 10^{-2} $ \\
0.232 & $ 1.24\times 10^{-2} $ \\
0.29 & $ 8.94\times 10^{-3} $ \\
0.348 & $ 6.52\times 10^{-3} $ \\
0.406 & $ 4.75\times 10^{-3} $ \\
0.464 & $ 3.43\times 10^{-3} $ \\
0.522 & $ 2.43\times 10^{-3} $ \\
0.58 & $ 1.69\times 10^{-3} $ \\
0.638 & $ 1.15\times 10^{-3} $ \\
0.697 & $ 7.75\times 10^{-4} $ \\
0.755 & $ 5.28\times 10^{-4} $ \\
0.813 & $ 3.69\times 10^{-4} $ \\
0.871 & $ 2.64\times 10^{-4} $ \\
0.929 & $ 1.94\times 10^{-4} $ \\
0.987 & $ 1.45\times 10^{-4} $ \\
1.04 & $ 1.10\times 10^{-4} $ \\
1.1 & $ 8.36\times 10^{-5} $ \\
1.16 & $ 6.36\times 10^{-5} $ \\
1.22 & $ 4.82\times 10^{-5} $ \\
1.28 & $ 3.61\times 10^{-5} $ \\
1.34 & $ 2.68\times 10^{-5} $ \\
1.39 & $ 1.95\times 10^{-5} $ \\
1.45 & $ 1.40\times 10^{-5} $ \\
1.51 & $ 9.92\times 10^{-6} $ \\
1.57 & $ 6.94\times 10^{-6} $ \\
1.63 & $ 4.86\times 10^{-6} $ \\
1.68 & $ 3.43\times 10^{-6} $ \\
1.74 & $ 2.47\times 10^{-6} $ \\
\end{tabular}
\begin{tabular}{c|c}
1.8 & $ 1.81\times 10^{-6} $ \\
1.86 & $ 1.35\times 10^{-6} $ \\
1.92 & $ 1.02\times 10^{-6} $ \\
1.97 & $ 7.76\times 10^{-7} $ \\
2.03 & $ 5.95\times 10^{-7} $ \\
2.09 & $ 4.57\times 10^{-7} $ \\
2.15 & $ 3.50\times 10^{-7} $ \\
2.21 & $ 2.67\times 10^{-7} $ \\
2.26 & $ 2.02\times 10^{-7} $ \\
2.32 & $ 1.51\times 10^{-7} $ \\
2.38 & $ 1.12\times 10^{-7} $ \\
2.44 & $ 8.15\times 10^{-8} $ \\
2.5 & $ 5.84\times 10^{-8} $ \\
2.55 & $ 4.13\times 10^{-8} $ \\
2.61 & $ 2.89\times 10^{-8} $ \\
2.67 & $ 2.02\times 10^{-8} $ \\
2.73 & $ 1.42\times 10^{-8} $ \\
2.79 & $ 1.02\times 10^{-8} $ \\
2.84 & $ 7.45\times 10^{-9} $ \\
2.9 & $ 5.54\times 10^{-9} $ \\
2.96 & $ 4.18\times 10^{-9} $ \\
3.02 & $ 3.19\times 10^{-9} $ \\
3.08 & $ 2.45\times 10^{-9} $ \\
3.13 & $ 1.89\times 10^{-9} $ \\
3.19 & $ 1.46\times 10^{-9} $ \\
3.25 & $ 1.12\times 10^{-9} $ \\
3.31 & $ 8.56\times 10^{-10} $ \\
3.37 & $ 6.51\times 10^{-10} $ \\
3.42 & $ 4.90\times 10^{-10} $ \\
3.48 & $ 3.66\times 10^{-10} $ \\
3.54 & $ 2.70\times 10^{-10} $ \\
\end{tabular}
\begin{tabular}{c|c}
3.6 & $ 1.96\times 10^{-10} $ \\
3.66 & $ 1.41\times 10^{-10} $ \\
3.71 & $ 9.99\times 10^{-11} $ \\
3.77 & $ 6.99\times 10^{-11} $ \\
3.83 & $ 4.85\times 10^{-11} $ \\
3.89 & $ 3.37\times 10^{-11} $ \\
3.95 & $ 2.37\times 10^{-11} $ \\
4.01 & $ 1.69\times 10^{-11} $ \\
4.06 & $ 1.23\times 10^{-11} $ \\
4.12 & $ 9.14\times 10^{-12} $ \\
4.18 & $ 6.88\times 10^{-12} $ \\
4.24 & $ 5.23\times 10^{-12} $ \\
4.3 & $ 3.99\times 10^{-12} $ \\
4.35 & $ 3.05\times 10^{-12} $ \\
4.41 & $ 2.31\times 10^{-12} $ \\
4.47 & $ 1.74\times 10^{-12} $ \\
4.53 & $ 1.30\times 10^{-12} $ \\
4.59 & $ 9.59\times 10^{-13} $ \\
4.64 & $ 7.02\times 10^{-13} $ \\
4.7 & $ 5.12\times 10^{-13} $ \\
4.76 & $ 3.73\times 10^{-13} $ \\
4.82 & $ 2.73\times 10^{-13} $ \\
4.88 & $ 1.99\times 10^{-13} $ \\
4.93 & $ 1.46\times 10^{-13} $ \\
4.99 & $ 1.06\times 10^{-13} $ \\
5.05 & $ 7.65\times 10^{-14} $ \\
5.11 & $ 5.46\times 10^{-14} $ \\
5.17 & $ 3.87\times 10^{-14} $ \\
5.22 & $ 2.74\times 10^{-14} $ \\
5.28 & $ 1.94\times 10^{-14} $ \\
5.34 & $ 1.39\times 10^{-14} $ \\
\end{tabular}
\begin{tabular}{c|c}
5.4 & $ 1.02\times 10^{-14} $ \\
5.46 & $ 7.53\times 10^{-15} $ \\
5.51 & $ 5.67\times 10^{-15} $ \\
5.57 & $ 4.32\times 10^{-15} $ \\
5.63 & $ 3.32\times 10^{-15} $ \\
5.69 & $ 2.56\times 10^{-15} $ \\
5.75 & $ 1.97\times 10^{-15} $ \\
5.8 & $ 1.52\times 10^{-15} $ \\
5.86 & $ 1.16\times 10^{-15} $ \\
5.92 & $ 8.77\times 10^{-16} $ \\
5.98 & $ 6.57\times 10^{-16} $ \\
6.04 & $ 4.86\times 10^{-16} $ \\
6.09 & $ 3.56\times 10^{-16} $ \\
6.15 & $ 2.58\times 10^{-16} $ \\
6.21 & $ 1.85\times 10^{-16} $ \\
6.27 & $ 1.32\times 10^{-16} $ \\
6.33 & $ 9.31\times 10^{-17} $ \\
6.38 & $ 6.51\times 10^{-17} $ \\
6.44 & $ 4.50\times 10^{-17} $ \\
6.5 & $ 3.08\times 10^{-17} $ \\
6.56 & $ 2.08\times 10^{-17} $ \\
6.62 & $ 1.40\times 10^{-17} $ \\
6.68 & $ 9.37\times 10^{-18} $ \\
6.73 & $ 6.28\times 10^{-18} $ \\
6.79 & $ 4.23\times 10^{-18} $ \\
6.85 & $ 2.85\times 10^{-18} $ \\
6.91 & $ 1.94\times 10^{-18} $ \\
6.97 & $ 1.32\times 10^{-18} $ \\
7.02 & $ 9.18\times 10^{-19} $ \\
7.08 & $ 6.47\times 10^{-19} $ \\
7.14 & $ 4.65\times 10^{-19} $ \\
\end{tabular}
\begin{tabular}{c|c}
7.2 & $ 3.41\times 10^{-19} $ \\
7.26 & $ 2.54\times 10^{-19} $ \\
7.31 & $ 1.91\times 10^{-19} $ \\
7.37 & $ 1.46\times 10^{-19} $ \\
7.43 & $ 1.11\times 10^{-19} $ \\
7.49 & $ 8.53\times 10^{-20} $ \\
7.55 & $ 6.50\times 10^{-20} $ \\
7.6 & $ 4.91\times 10^{-20} $ \\
7.66 & $ 3.66\times 10^{-20} $ \\
7.72 & $ 2.67\times 10^{-20} $ \\
7.78 & $ 1.89\times 10^{-20} $ \\
7.84 & $ 1.30\times 10^{-20} $ \\
7.89 & $ 8.61\times 10^{-21} $ \\
7.95 & $ 5.52\times 10^{-21} $ \\
8.01 & $ 3.44\times 10^{-21} $ \\
8.07 & $ 2.10\times 10^{-21} $ \\
8.13 & $ 1.27\times 10^{-21} $ \\
8.18 & $ 7.92\times 10^{-22} $ \\
8.24 & $ 5.16\times 10^{-22} $ \\
8.3 & $ 3.53\times 10^{-22} $ \\
8.36 & $ 2.50\times 10^{-22} $ \\
8.42 & $ 1.81\times 10^{-22} $ \\
8.47 & $ 1.33\times 10^{-22} $ \\
8.53 & $ 9.83\times 10^{-23} $ \\
8.59 & $ 7.24\times 10^{-23} $ \\
8.65 & $ 5.28\times 10^{-23} $ \\
8.71 & $ 3.79\times 10^{-23} $ \\
8.76 & $ 2.65\times 10^{-23} $ \\
8.82 & $ 1.80\times 10^{-23} $ \\
8.88 & $ 1.17\times 10^{-23} $ \\
8.94 & $ 7.27\times 10^{-24} $ \\
\hline
\hline
\end{tabular}
  \caption{Logarithmic negativity as a function of separation measured in units of the averaged region diameter with an N-type boundary appearing in the left panel of Fig.~\ref{fig:neg2d}.}
  \label{tab:2dnegativity4}
\end{table}

\begin{table}
\footnotesize
\begin{tabular}{c|c}
\multicolumn{2}{c}{$d_{lat} = 21$} \\
\hline
\hline
$\tilde{r}/d_{avg}$& $\mathcal{N}$ \\
\hline
\hline
2.68 & $ 1.72\times 10^{-8} $ \\
2.99 & $ 3.64\times 10^{-9} $ \\
3.3 & $ 8.75\times 10^{-10} $ \\
3.61 & $ 1.52\times 10^{-10} $ \\
3.92 & $ 2.60\times 10^{-11} $ \\
4.23 & $ 6.18\times 10^{-12} $ \\
4.54 & $ 1.33\times 10^{-12} $ \\
4.85 & $ 2.17\times 10^{-13} $ \\
5.16 & $ 3.24\times 10^{-14} $ \\
5.47 & $ 6.89\times 10^{-15} $ \\
5.78 & $ 1.50\times 10^{-15} $ \\
6.09 & $ 2.66\times 10^{-16} $ \\
6.4 & $ 3.76\times 10^{-17} $ \\
6.71 & $ 6.37\times 10^{-18} $ \\
7.02 & $ 1.56\times 10^{-18} $ \\
7.33 & $ 3.22\times 10^{-19} $ \\
7.64 & $ 4.47\times 10^{-20} $ \\
7.95 & $ 6.12\times 10^{-21} $ \\
8.26 & $ 9.22\times 10^{-22} $ \\
8.57 & $ 1.34\times 10^{-22} $ \\
\hline
\hline
\end{tabular}
\begin{tabular}{c|c}
\multicolumn{2}{c}{$d_{lat} = 23$} \\
\hline
\hline
$\tilde{r}/d_{avg}$& $\mathcal{N}$ \\
\hline
\hline
2.69 & $ 1.58\times 10^{-8} $ \\
2.97 & $ 4.29\times 10^{-9} $ \\
3.26 & $ 1.12\times 10^{-9} $ \\
3.54 & $ 2.21\times 10^{-10} $ \\
3.82 & $ 4.56\times 10^{-11} $ \\
4.11 & $ 1.18\times 10^{-11} $ \\
4.39 & $ 2.77\times 10^{-12} $ \\
4.67 & $ 5.72\times 10^{-13} $ \\
4.96 & $ 1.12\times 10^{-13} $ \\
5.24 & $ 2.91\times 10^{-14} $ \\
5.52 & $ 7.52\times 10^{-15} $ \\
5.81 & $ 1.44\times 10^{-15} $ \\
6.09 & $ 2.53\times 10^{-16} $ \\
6.37 & $ 5.22\times 10^{-17} $ \\
6.66 & $ 1.37\times 10^{-17} $ \\
6.94 & $ 2.87\times 10^{-18} $ \\
7.22 & $ 4.72\times 10^{-19} $ \\
7.51 & $ 8.00\times 10^{-20} $ \\
7.79 & $ 1.60\times 10^{-20} $ \\
8.07 & $ 3.96\times 10^{-21} $ \\
8.35 & $ 9.64\times 10^{-22} $ \\
8.64 & $ 1.70\times 10^{-22} $ \\
\hline
\hline
\end{tabular}
\begin{tabular}{c|c}
\multicolumn{2}{c}{$d_{lat} = 25$} \\
\hline
\hline
$\tilde{r}/d_{avg}$& $\mathcal{N}$ \\
\hline
\hline
2.7 & $ 1.53\times 10^{-8} $ \\
2.96 & $ 4.59\times 10^{-9} $ \\
3.22 & $ 1.20\times 10^{-9} $ \\
3.48 & $ 2.61\times 10^{-10} $ \\
3.74 & $ 6.64\times 10^{-11} $ \\
4.01 & $ 1.96\times 10^{-11} $ \\
4.27 & $ 5.25\times 10^{-12} $ \\
4.53 & $ 1.19\times 10^{-12} $ \\
4.79 & $ 2.65\times 10^{-13} $ \\
5.05 & $ 7.62\times 10^{-14} $ \\
5.31 & $ 2.18\times 10^{-14} $ \\
5.57 & $ 5.10\times 10^{-15} $ \\
5.84 & $ 1.15\times 10^{-15} $ \\
6.1 & $ 2.63\times 10^{-16} $ \\
6.36 & $ 6.85\times 10^{-17} $ \\
6.62 & $ 1.89\times 10^{-17} $ \\
6.88 & $ 4.86\times 10^{-18} $ \\
7.14 & $ 1.10\times 10^{-18} $ \\
7.4 & $ 2.44\times 10^{-19} $ \\
7.66 & $ 5.84\times 10^{-20} $ \\
7.93 & $ 1.46\times 10^{-20} $ \\
8.19 & $ 3.36\times 10^{-21} $ \\
8.45 & $ 7.29\times 10^{-22} $ \\
8.71 & $ 1.53\times 10^{-22} $ \\
\hline
\hline
\end{tabular}
\begin{tabular}{c|c}
\multicolumn{2}{c}{$d_{lat} = 27$} \\
\hline
\hline
$\tilde{r}/d_{avg}$& $\mathcal{N}$ \\
\hline
\hline
2.66 & $ 2.07\times 10^{-8} $ \\
2.89 & $ 6.73\times 10^{-9} $ \\
3.13 & $ 1.98\times 10^{-9} $ \\
3.37 & $ 5.11\times 10^{-10} $ \\
3.61 & $ 1.39\times 10^{-10} $ \\
3.85 & $ 4.44\times 10^{-11} $ \\
4.08 & $ 1.39\times 10^{-11} $ \\
4.32 & $ 3.81\times 10^{-12} $ \\
4.56 & $ 9.90\times 10^{-13} $ \\
4.8 & $ 2.92\times 10^{-13} $ \\
5.04 & $ 9.12\times 10^{-14} $ \\
5.27 & $ 2.64\times 10^{-14} $ \\
5.51 & $ 7.18\times 10^{-15} $ \\
5.75 & $ 1.89\times 10^{-15} $ \\
5.99 & $ 5.22\times 10^{-16} $ \\
6.23 & $ 1.51\times 10^{-16} $ \\
6.46 & $ 3.98\times 10^{-17} $ \\
6.7 & $ 1.09\times 10^{-17} $ \\
6.94 & $ 3.02\times 10^{-18} $ \\
7.18 & $ 7.98\times 10^{-19} $ \\
7.41 & $ 2.28\times 10^{-19} $ \\
7.65 & $ 7.19\times 10^{-20} $ \\
7.89 & $ 2.20\times 10^{-20} $ \\
8.13 & $ 5.56\times 10^{-21} $ \\
8.37 & $ 1.28\times 10^{-21} $ \\
\hline
\hline
\end{tabular}
\begin{tabular}{c|c}
\multicolumn{2}{c}{$d_{lat} = 29$} \\
\hline
\hline
$\tilde{r}/d_{avg}$& $\mathcal{N}$ \\
\hline
\hline
2.65 & $ 2.30\times 10^{-8} $ \\
2.87 & $ 8.13\times 10^{-9} $ \\
3.09 & $ 2.53\times 10^{-9} $ \\
3.31 & $ 6.95\times 10^{-10} $ \\
3.53 & $ 2.14\times 10^{-10} $ \\
3.75 & $ 7.51\times 10^{-11} $ \\
3.97 & $ 2.55\times 10^{-11} $ \\
4.19 & $ 7.75\times 10^{-12} $ \\
4.41 & $ 2.2\times 10^{-12} $ \\
4.63 & $ 6.91\times 10^{-13} $ \\
4.85 & $ 2.35\times 10^{-13} $ \\
5.07 & $ 7.55\times 10^{-14} $ \\
5.29 & $ 2.27\times 10^{-14} $ \\
5.51 & $ 6.62\times 10^{-15} $ \\
5.74 & $ 2.\times 10^{-15} $ \\
5.96 & $ 6.81\times 10^{-16} $ \\
6.18 & $ 2.29\times 10^{-16} $ \\
6.4 & $ 6.81\times 10^{-17} $ \\
6.62 & $ 1.99\times 10^{-17} $ \\
6.84 & $ 5.95\times 10^{-18} $ \\
7.06 & $ 1.79\times 10^{-18} $ \\
7.28 & $ 5.68\times 10^{-19} $ \\
7.5 & $ 1.82\times 10^{-19} $ \\
7.72 & $ 5.8\times 10^{-20} $ \\
7.94 & $ 1.87\times 10^{-20} $ \\
8.16 & $ 5.68\times 10^{-21} $ \\
8.38 & $ 1.55\times 10^{-21} $ \\
\hline
\hline
\end{tabular}
  \caption{Logarithmic negativity as a function of separation measured in units of the averaged region diameter with an N-type boundary appearing in the left panel of Fig.~\ref{fig:neg2d}.}
  \label{tab:2dnegativity5}
\end{table}

\begin{table}
\footnotesize
\begin{tabular}{c|c}
\multicolumn{2}{c}{$d_{lat} = 31$} \\
\hline
\hline
$\tilde{r}/d_{avg}$& $\mathcal{N}$ \\
\hline
\hline
2.63 & $ 2.66\times 10^{-8} $ \\
2.8 & $ 1.19\times 10^{-8} $ \\
2.97 & $ 4.92\times 10^{-9} $ \\
3.14 & $ 1.82\times 10^{-9} $ \\
3.31 & $ 6.67\times 10^{-10} $ \\
3.48 & $ 2.73\times 10^{-10} $ \\
3.65 & $ 1.22\times 10^{-10} $ \\
3.82 & $ 5.47\times 10^{-11} $ \\
3.99 & $ 2.28\times 10^{-11} $ \\
4.16 & $ 8.64\times 10^{-12} $ \\
4.33 & $ 3.22\times 10^{-12} $ \\
4.5 & $ 1.33\times 10^{-12} $ \\
4.67 & $ 5.97\times 10^{-13} $ \\
4.84 & $ 2.64\times 10^{-13} $ \\
5.01 & $ 1.08\times 10^{-13} $ \\
5.18 & $ 4.11\times 10^{-14} $ \\
5.36 & $ 1.49\times 10^{-14} $ \\
5.53 & $ 5.50\times 10^{-15} $ \\
5.7 & $ 2.29\times 10^{-15} $ \\
5.87 & $ 1.02\times 10^{-15} $ \\
6.04 & $ 4.45\times 10^{-16} $ \\
6.21 & $ 1.77\times 10^{-16} $ \\
6.38 & $ 6.85\times 10^{-17} $ \\
6.55 & $ 2.70\times 10^{-17} $ \\
6.72 & $ 1.08\times 10^{-17} $ \\
6.89 & $ 4.44\times 10^{-18} $ \\
7.06 & $ 1.89\times 10^{-18} $ \\
7.23 & $ 7.85\times 10^{-19} $ \\
7.4 & $ 2.98\times 10^{-19} $ \\
7.57 & $ 1.06\times 10^{-19} $ \\
7.74 & $ 3.91\times 10^{-20} $ \\
7.91 & $ 1.51\times 10^{-20} $ \\
8.08 & $ 6.02\times 10^{-21} $ \\
8.25 & $ 2.53\times 10^{-21} $ \\
8.43 & $ 1.10\times 10^{-21} $ \\
\hline
\hline
\end{tabular}
\begin{tabular}{c|c}
\multicolumn{2}{c}{$d_{lat} = 33$} \\
\hline
\hline
$\tilde{r}/d_{avg}$& $\mathcal{N}$ \\
\hline
\hline
2.64 & $ 2.55\times 10^{-8} $ \\
2.8 & $ 1.15\times 10^{-8} $ \\
2.96 & $ 4.78\times 10^{-9} $ \\
3.12 & $ 1.85\times 10^{-9} $ \\
3.28 & $ 7.49\times 10^{-10} $ \\
3.44 & $ 3.36\times 10^{-10} $ \\
3.6 & $ 1.58\times 10^{-10} $ \\
3.76 & $ 7.28\times 10^{-11} $ \\
3.92 & $ 3.10\times 10^{-11} $ \\
4.08 & $ 1.23\times 10^{-11} $ \\
4.24 & $ 4.93\times 10^{-12} $ \\
4.4 & $ 2.17\times 10^{-12} $ \\
4.56 & $ 1.02\times 10^{-12} $ \\
4.72 & $ 4.77\times 10^{-13} $ \\
4.88 & $ 2.10\times 10^{-13} $ \\
5.05 & $ 8.69\times 10^{-14} $ \\
5.21 & $ 3.42\times 10^{-14} $ \\
5.37 & $ 1.34\times 10^{-14} $ \\
5.53 & $ 5.67\times 10^{-15} $ \\
5.69 & $ 2.62\times 10^{-15} $ \\
5.85 & $ 1.22\times 10^{-15} $ \\
6.01 & $ 5.39\times 10^{-16} $ \\
6.17 & $ 2.25\times 10^{-16} $ \\
6.33 & $ 9.31\times 10^{-17} $ \\
6.49 & $ 3.91\times 10^{-17} $ \\
6.65 & $ 1.67\times 10^{-17} $ \\
6.81 & $ 7.22\times 10^{-18} $ \\
6.97 & $ 3.12\times 10^{-18} $ \\
7.13 & $ 1.32\times 10^{-18} $ \\
7.29 & $ 5.53\times 10^{-19} $ \\
7.46 & $ 2.27\times 10^{-19} $ \\
7.62 & $ 9.33\times 10^{-20} $ \\
7.78 & $ 3.93\times 10^{-20} $ \\
7.94 & $ 1.69\times 10^{-20} $ \\
8.1 & $ 7.37\times 10^{-21} $ \\
8.26 & $ 3.14\times 10^{-21} $ \\
8.42 & $ 1.28\times 10^{-21} $ \\
\hline
\hline
\end{tabular}
\begin{tabular}{c|c}
\multicolumn{2}{c}{$d_{lat} = 35$} \\
\hline
\hline
$\tilde{r}/d_{avg}$& $\mathcal{N}$ \\
\hline
\hline
2.63 & $ 2.64\times 10^{-8} $ \\
2.81 & $ 1.06\times 10^{-8} $ \\
2.99 & $ 3.85\times 10^{-9} $ \\
3.17 & $ 1.36\times 10^{-9} $ \\
3.35 & $ 5.30\times 10^{-10} $ \\
3.53 & $ 2.21\times 10^{-10} $ \\
3.71 & $ 9.00\times 10^{-11} $ \\
3.89 & $ 3.41\times 10^{-11} $ \\
4.07 & $ 1.21\times 10^{-11} $ \\
4.26 & $ 4.44\times 10^{-12} $ \\
4.44 & $ 1.81\times 10^{-12} $ \\
4.62 & $ 7.51\times 10^{-13} $ \\
4.8 & $ 2.98\times 10^{-13} $ \\
4.98 & $ 1.13\times 10^{-13} $ \\
5.16 & $ 4.10\times 10^{-14} $ \\
5.34 & $ 1.53\times 10^{-14} $ \\
5.52 & $ 6.01\times 10^{-15} $ \\
5.7 & $ 2.41\times 10^{-15} $ \\
5.89 & $ 9.52\times 10^{-16} $ \\
6.07 & $ 3.73\times 10^{-16} $ \\
6.25 & $ 1.41\times 10^{-16} $ \\
6.43 & $ 5.17\times 10^{-17} $ \\
6.61 & $ 1.97\times 10^{-17} $ \\
6.79 & $ 7.85\times 10^{-18} $ \\
6.97 & $ 3.07\times 10^{-18} $ \\
7.15 & $ 1.18\times 10^{-18} $ \\
7.33 & $ 4.54\times 10^{-19} $ \\
7.52 & $ 1.69\times 10^{-19} $ \\
7.7 & $ 5.99\times 10^{-20} $ \\
7.88 & $ 2.17\times 10^{-20} $ \\
8.06 & $ 8.33\times 10^{-21} $ \\
8.24 & $ 3.37\times 10^{-21} $ \\
8.42 & $ 1.40\times 10^{-21} $ \\
\hline
\hline
\end{tabular}
\begin{tabular}{c|c}
\multicolumn{2}{c}{$d_{lat} = 37$} \\
\hline
\hline
$\tilde{r}/d_{avg}$& $\mathcal{N}$ \\
\hline
\hline
2.62 & $ 2.79\times 10^{-8} $ \\
2.79 & $ 1.16\times 10^{-8} $ \\
2.96 & $ 4.39\times 10^{-9} $ \\
3.13 & $ 1.65\times 10^{-9} $ \\
3.3 & $ 6.91\times 10^{-10} $ \\
3.48 & $ 3.04\times 10^{-10} $ \\
3.65 & $ 1.29\times 10^{-10} $ \\
3.82 & $ 5.09\times 10^{-11} $ \\
3.99 & $ 1.91\times 10^{-11} $ \\
4.16 & $ 7.45\times 10^{-12} $ \\
4.33 & $ 3.14\times 10^{-12} $ \\
4.5 & $ 1.38\times 10^{-12} $ \\
4.67 & $ 5.96\times 10^{-13} $ \\
4.84 & $ 2.46\times 10^{-13} $ \\
5.01 & $ 9.52\times 10^{-14} $ \\
5.18 & $ 3.67\times 10^{-14} $ \\
5.36 & $ 1.52\times 10^{-14} $ \\
5.53 & $ 6.57\times 10^{-15} $ \\
5.7 & $ 2.76\times 10^{-15} $ \\
5.87 & $ 1.12\times 10^{-15} $ \\
6.04 & $ 4.50\times 10^{-16} $ \\
6.21 & $ 1.77\times 10^{-16} $ \\
6.38 & $ 6.96\times 10^{-17} $ \\
6.55 & $ 2.84\times 10^{-17} $ \\
6.72 & $ 1.19\times 10^{-17} $ \\
6.89 & $ 4.93\times 10^{-18} $ \\
7.06 & $ 2.02\times 10^{-18} $ \\
7.24 & $ 8.30\times 10^{-19} $ \\
7.41 & $ 3.34\times 10^{-19} $ \\
7.58 & $ 1.30\times 10^{-19} $ \\
7.75 & $ 5.14\times 10^{-20} $ \\
7.92 & $ 2.13\times 10^{-20} $ \\
8.09 & $ 9.06\times 10^{-21} $ \\
8.26 & $ 3.84\times 10^{-21} $ \\
8.43 & $ 1.57\times 10^{-21} $ \\
\hline
\hline
\end{tabular}
\begin{tabular}{c|c}
\multicolumn{2}{c}{$d_{lat} = 39$} \\
\hline
\hline
$\tilde{r}/d_{avg}$& $\mathcal{N}$ \\
\hline
\hline
2.61 & $ 3.01\times 10^{-8} $ \\
2.77 & $ 1.26\times 10^{-8} $ \\
2.93 & $ 4.92\times 10^{-9} $ \\
3.09 & $ 2.00\times 10^{-9} $ \\
3.26 & $ 8.91\times 10^{-10} $ \\
3.42 & $ 4.13\times 10^{-10} $ \\
3.58 & $ 1.85\times 10^{-10} $ \\
3.74 & $ 7.77\times 10^{-11} $ \\
3.9 & $ 3.11\times 10^{-11} $ \\
4.06 & $ 1.29\times 10^{-11} $ \\
4.22 & $ 5.75\times 10^{-12} $ \\
4.39 & $ 2.64\times 10^{-12} $ \\
4.55 & $ 1.18\times 10^{-12} $ \\
4.71 & $ 4.98\times 10^{-13} $ \\
4.87 & $ 2.03\times 10^{-13} $ \\
5.03 & $ 8.17\times 10^{-14} $ \\
5.19 & $ 3.46\times 10^{-14} $ \\
5.35 & $ 1.55\times 10^{-14} $ \\
5.52 & $ 7.04\times 10^{-15} $ \\
5.68 & $ 3.08\times 10^{-15} $ \\
5.84 & $ 1.31\times 10^{-15} $ \\
6. & $ 5.45\times 10^{-16} $ \\
6.16 & $ 2.28\times 10^{-16} $ \\
6.32 & $ 9.80\times 10^{-17} $ \\
6.48 & $ 4.32\times 10^{-17} $ \\
6.65 & $ 1.88\times 10^{-17} $ \\
6.81 & $ 7.94\times 10^{-18} $ \\
6.97 & $ 3.34\times 10^{-18} $ \\
7.13 & $ 1.42\times 10^{-18} $ \\
7.29 & $ 5.97\times 10^{-19} $ \\
7.45 & $ 2.48\times 10^{-19} $ \\
7.61 & $ 1.04\times 10^{-19} $ \\
7.78 & $ 4.40\times 10^{-20} $ \\
7.94 & $ 1.87\times 10^{-20} $ \\
8.1 & $ 8.05\times 10^{-21} $ \\
8.26 & $ 3.48\times 10^{-21} $ \\
\hline
\hline
\end{tabular}
  \caption{Logarithmic negativity as a function of separation measured in units of the averaged region diameter with an N-type boundary appearing in the left panel of Fig.~\ref{fig:neg2d}.}
  \label{tab:2dnegativity6}
\end{table}

\begin{table}
\begin{tabular}{c|c}
\multicolumn{2}{c}{$d_{lat} = 41$} \\
\hline
\hline
$\tilde{r}/d_{avg}$& $\mathcal{N}$ \\
\hline
\hline
2.6 & $ 3.24\times 10^{-8} $ \\
2.8 & $ 1.04\times 10^{-8} $ \\
3.01 & $ 3.20\times 10^{-9} $ \\
3.21 & $ 1.13\times 10^{-9} $ \\
3.41 & $ 4.28\times 10^{-10} $ \\
3.62 & $ 1.51\times 10^{-10} $ \\
3.82 & $ 4.76\times 10^{-11} $ \\
4.02 & $ 1.55\times 10^{-11} $ \\
4.23 & $ 5.74\times 10^{-12} $ \\
4.43 & $ 2.16\times 10^{-12} $ \\
4.63 & $ 7.38\times 10^{-13} $ \\
4.84 & $ 2.30\times 10^{-13} $ \\
5.04 & $ 7.40\times 10^{-14} $ \\
5.25 & $ 2.71\times 10^{-14} $ \\
5.45 & $ 1.02\times 10^{-14} $ \\
5.65 & $ 3.52\times 10^{-15} $ \\
5.86 & $ 1.14\times 10^{-15} $ \\
6.06 & $ 3.67\times 10^{-16} $ \\
6.26 & $ 1.27\times 10^{-16} $ \\
6.47 & $ 4.65\times 10^{-17} $ \\
6.67 & $ 1.63\times 10^{-17} $ \\
6.88 & $ 5.37\times 10^{-18} $ \\
7.08 & $ 1.78\times 10^{-18} $ \\
7.28 & $ 6.05\times 10^{-19} $ \\
7.49 & $ 2.09\times 10^{-19} $ \\
7.69 & $ 7.12\times 10^{-20} $ \\
7.89 & $ 2.33\times 10^{-20} $ \\
8.1 & $ 7.48\times 10^{-21} $ \\
8.3 & $ 2.54\times 10^{-21} $ \\
\hline
\hline
\end{tabular}
\begin{tabular}{c|c}
\multicolumn{2}{c}{$d_{lat} = 43$} \\
\hline
\hline
$\tilde{r}/d_{avg}$& $\mathcal{N}$ \\
\hline
\hline
2.61 & $ 3.00\times 10^{-8} $ \\
2.8 & $ 9.93\times 10^{-9} $ \\
3. & $ 3.29\times 10^{-9} $ \\
3.19 & $ 1.25\times 10^{-9} $ \\
3.39 & $ 4.92\times 10^{-10} $ \\
3.58 & $ 1.80\times 10^{-10} $ \\
3.77 & $ 5.94\times 10^{-11} $ \\
3.97 & $ 2.04\times 10^{-11} $ \\
4.16 & $ 7.83\times 10^{-12} $ \\
4.36 & $ 3.04\times 10^{-12} $ \\
4.55 & $ 1.09\times 10^{-12} $ \\
4.75 & $ 3.64\times 10^{-13} $ \\
4.94 & $ 1.24\times 10^{-13} $ \\
5.14 & $ 4.65\times 10^{-14} $ \\
5.33 & $ 1.79\times 10^{-14} $ \\
5.53 & $ 6.46\times 10^{-15} $ \\
5.72 & $ 2.21\times 10^{-15} $ \\
5.92 & $ 7.55\times 10^{-16} $ \\
6.11 & $ 2.75\times 10^{-16} $ \\
6.31 & $ 1.06\times 10^{-16} $ \\
6.5 & $ 3.96\times 10^{-17} $ \\
6.7 & $ 1.36\times 10^{-17} $ \\
6.89 & $ 4.68\times 10^{-18} $ \\
7.09 & $ 1.68\times 10^{-18} $ \\
7.28 & $ 6.17\times 10^{-19} $ \\
7.48 & $ 2.30\times 10^{-19} $ \\
7.67 & $ 8.35\times 10^{-20} $ \\
7.87 & $ 2.90\times 10^{-20} $ \\
8.06 & $ 9.85\times 10^{-21} $ \\
8.26 & $ 3.46\times 10^{-21} $ \\
\hline
\hline
\end{tabular}
\begin{tabular}{c|c}
\multicolumn{2}{c}{$d_{lat} = 47$} \\
\hline
\hline
$\tilde{r}/d_{avg}$& $\mathcal{N}$ \\
\hline
\hline
2.59 & $ 3.16\times 10^{-8} $ \\
2.81 & $ 8.88\times 10^{-9} $ \\
3.04 & $ 2.74\times 10^{-9} $ \\
3.26 & $ 9.32\times 10^{-10} $ \\
3.48 & $ 2.98\times 10^{-10} $ \\
3.7 & $ 8.64\times 10^{-11} $ \\
3.92 & $ 2.65\times 10^{-11} $ \\
4.14 & $ 9.03\times 10^{-12} $ \\
4.37 & $ 2.93\times 10^{-12} $ \\
4.59 & $ 8.58\times 10^{-13} $ \\
4.81 & $ 2.54\times 10^{-13} $ \\
5.03 & $ 8.23\times 10^{-14} $ \\
5.25 & $ 2.73\times 10^{-14} $ \\
5.47 & $ 8.65\times 10^{-15} $ \\
5.69 & $ 2.56\times 10^{-15} $ \\
5.92 & $ 7.69\times 10^{-16} $ \\
6.14 & $ 2.50\times 10^{-16} $ \\
6.36 & $ 7.96\times 10^{-17} $ \\
6.58 & $ 2.50\times 10^{-17} $ \\
6.8 & $ 7.72\times 10^{-18} $ \\
7.02 & $ 2.33\times 10^{-18} $ \\
7.25 & $ 7.39\times 10^{-19} $ \\
7.47 & $ 2.44\times 10^{-19} $ \\
7.69 & $ 7.77\times 10^{-20} $ \\
7.91 & $ 2.37\times 10^{-20} $ \\
8.13 & $ 7.10\times 10^{-21} $ \\
\hline
\hline
\end{tabular}
\begin{tabular}{c|c}
\multicolumn{2}{c}{$d_{lat} = 49$} \\
\hline
\hline
$\tilde{r}/d_{avg}$& $\mathcal{N}$ \\
\hline
\hline
2.59 & $ 3.05\times 10^{-8} $ \\
2.76 & $ 1.15\times 10^{-8} $ \\
2.93 & $ 4.57\times 10^{-9} $ \\
3.1 & $ 1.98\times 10^{-9} $ \\
3.27 & $ 8.64\times 10^{-10} $ \\
3.44 & $ 3.54\times 10^{-10} $ \\
3.61 & $ 1.36\times 10^{-10} $ \\
3.78 & $ 5.31\times 10^{-11} $ \\
3.96 & $ 2.25\times 10^{-11} $ \\
4.13 & $ 9.87\times 10^{-12} $ \\
4.3 & $ 4.16\times 10^{-12} $ \\
4.47 & $ 1.65\times 10^{-12} $ \\
4.64 & $ 6.37\times 10^{-13} $ \\
4.81 & $ 2.58\times 10^{-13} $ \\
4.98 & $ 1.11\times 10^{-13} $ \\
5.15 & $ 4.75\times 10^{-14} $ \\
5.32 & $ 1.95\times 10^{-14} $ \\
5.49 & $ 7.73\times 10^{-15} $ \\
5.66 & $ 2.99\times 10^{-15} $ \\
5.83 & $ 1.21\times 10^{-15} $ \\
6. & $ 5.17\times 10^{-16} $ \\
6.17 & $ 2.22\times 10^{-16} $ \\
6.34 & $ 9.13\times 10^{-17} $ \\
6.51 & $ 3.66\times 10^{-17} $ \\
6.68 & $ 1.47\times 10^{-17} $ \\
6.85 & $ 5.99\times 10^{-18} $ \\
7.02 & $ 2.49\times 10^{-18} $ \\
7.19 & $ 1.05\times 10^{-18} $ \\
7.36 & $ 4.35\times 10^{-19} $ \\
7.53 & $ 1.79\times 10^{-19} $ \\
7.7 & $ 7.23\times 10^{-20} $ \\
7.87 & $ 2.86\times 10^{-20} $ \\
8.04 & $ 1.14\times 10^{-20} $ \\
8.21 & $ 4.69\times 10^{-21} $ \\
\hline
\hline
\end{tabular}
  \caption{Logarithmic negativity as a function of separation measured in units of the averaged region diameter with an N-type boundary appearing in the left panel of Fig.~\ref{fig:neg2d}.}
  \label{tab:2dnegativity7}
\end{table}

\begin{table}
\begin{tabular}{c|c}
\multicolumn{2}{c}{$d_{lat} = 57$} \\
\hline
\hline
$\tilde{r}/d_{avg}$& $\mathcal{N}$ \\
\hline
\hline
2.58 & $ 3.22\times 10^{-8} $ \\
2.76 & $ 1.17\times 10^{-8} $ \\
2.94 & $ 4.71\times 10^{-9} $ \\
3.12 & $ 1.93\times 10^{-9} $ \\
3.3 & $ 7.43\times 10^{-10} $ \\
3.48 & $ 2.68\times 10^{-10} $ \\
3.66 & $ 1.01\times 10^{-10} $ \\
3.85 & $ 4.09\times 10^{-11} $ \\
4.03 & $ 1.66\times 10^{-11} $ \\
4.21 & $ 6.31\times 10^{-12} $ \\
4.39 & $ 2.30\times 10^{-12} $ \\
4.57 & $ 8.73\times 10^{-13} $ \\
4.75 & $ 3.52\times 10^{-13} $ \\
4.93 & $ 1.42\times 10^{-13} $ \\
5.11 & $ 5.45\times 10^{-14} $ \\
5.3 & $ 2.02\times 10^{-14} $ \\
5.48 & $ 7.49\times 10^{-15} $ \\
5.66 & $ 2.96\times 10^{-15} $ \\
5.84 & $ 1.21\times 10^{-15} $ \\
6.02 & $ 4.81\times 10^{-16} $ \\
6.2 & $ 1.84\times 10^{-16} $ \\
6.38 & $ 6.89\times 10^{-17} $ \\
6.57 & $ 2.61\times 10^{-17} $ \\
6.75 & $ 1.02\times 10^{-17} $ \\
6.93 & $ 4.01\times 10^{-18} $ \\
7.11 & $ 1.55\times 10^{-18} $ \\
7.29 & $ 5.96\times 10^{-19} $ \\
7.47 & $ 2.25\times 10^{-19} $ \\
7.65 & $ 8.42\times 10^{-20} $ \\
7.84 & $ 3.25\times 10^{-20} $ \\
8.02 & $ 1.29\times 10^{-20} $ \\
8.2 & $ 5.11\times 10^{-21} $ \\
\hline
\hline
\end{tabular}
\begin{tabular}{c|c}
\multicolumn{2}{c}{$d_{lat} = 67$} \\
\hline
\hline
$\tilde{r}/d_{avg}$& $\mathcal{N}$ \\
\hline
\hline
2.56 & $ 3.28\times 10^{-8} $ \\
2.75 & $ 1.26\times 10^{-8} $ \\
2.93 & $ 5.04\times 10^{-9} $ \\
3.12 & $ 1.93\times 10^{-9} $ \\
3.3 & $ 6.89\times 10^{-10} $ \\
3.49 & $ 2.54\times 10^{-10} $ \\
3.67 & $ 1.01\times 10^{-10} $ \\
3.85 & $ 4.00\times 10^{-11} $ \\
4.04 & $ 1.49\times 10^{-11} $ \\
4.22 & $ 5.37\times 10^{-12} $ \\
4.41 & $ 2.05\times 10^{-12} $ \\
4.59 & $ 8.22\times 10^{-13} $ \\
4.78 & $ 3.21\times 10^{-13} $ \\
4.96 & $ 1.18\times 10^{-13} $ \\
5.14 & $ 4.27\times 10^{-14} $ \\
5.33 & $ 1.63\times 10^{-14} $ \\
5.51 & $ 6.49\times 10^{-15} $ \\
5.7 & $ 2.55\times 10^{-15} $ \\
5.88 & $ 9.57\times 10^{-16} $ \\
6.07 & $ 3.51\times 10^{-16} $ \\
6.25 & $ 1.32\times 10^{-16} $ \\
6.43 & $ 5.12\times 10^{-17} $ \\
6.62 & $ 1.99\times 10^{-17} $ \\
6.8 & $ 7.60\times 10^{-18} $ \\
6.99 & $ 2.87\times 10^{-18} $ \\
7.17 & $ 1.08\times 10^{-18} $ \\
7.36 & $ 4.13\times 10^{-19} $ \\
7.54 & $ 1.60\times 10^{-19} $ \\
7.72 & $ 6.14\times 10^{-20} $ \\
7.91 & $ 2.33\times 10^{-20} $ \\
8.09 & $ 8.66\times 10^{-21} $ \\
\hline
\hline
\end{tabular}
\begin{tabular}{c|c}
\multicolumn{2}{c}{$d_{lat} = 79$} \\
\hline
\hline
$\tilde{r}/d_{avg}$& $\mathcal{N}$ \\
\hline
\hline
2.55 & $ 3.54\times 10^{-8} $ \\
2.76 & $ 1.25\times 10^{-8} $ \\
2.96 & $ 4.26\times 10^{-9} $ \\
3.17 & $ 1.35\times 10^{-9} $ \\
3.38 & $ 4.46\times 10^{-10} $ \\
3.59 & $ 1.58\times 10^{-10} $ \\
3.79 & $ 5.38\times 10^{-11} $ \\
4. & $ 1.72\times 10^{-11} $ \\
4.21 & $ 5.75\times 10^{-12} $ \\
4.42 & $ 2.04\times 10^{-12} $ \\
4.62 & $ 7.00\times 10^{-13} $ \\
4.83 & $ 2.25\times 10^{-13} $ \\
5.04 & $ 7.37\times 10^{-14} $ \\
5.24 & $ 2.59\times 10^{-14} $ \\
5.45 & $ 8.97\times 10^{-15} $ \\
5.66 & $ 2.96\times 10^{-15} $ \\
5.87 & $ 9.68\times 10^{-16} $ \\
6.07 & $ 3.31\times 10^{-16} $ \\
6.28 & $ 1.15\times 10^{-16} $ \\
6.49 & $ 3.94\times 10^{-17} $ \\
6.69 & $ 1.31\times 10^{-17} $ \\
6.9 & $ 4.32\times 10^{-18} $ \\
7.11 & $ 1.48\times 10^{-18} $ \\
7.32 & $ 5.05\times 10^{-19} $ \\
7.52 & $ 1.71\times 10^{-19} $ \\
7.73 & $ 5.63\times 10^{-20} $ \\
7.94 & $ 1.87\times 10^{-20} $ \\
8.14 & $ 6.35\times 10^{-21} $ \\
\hline
\hline
\end{tabular}
  \caption{Logarithmic negativity as a function of separation measured in units of the averaged region diameter with an N-type boundary appearing in the left panel of Fig.~\ref{fig:neg2d}.}
  \label{tab:2dnegativity8}
\end{table}

\begin{table}
\begin{tabular}{c|c}
  \multicolumn{2}{c}{$N$ boundary} \\
  \hline
  \hline
  $1/d_{lat}$ & $\beta_{2D}$ \\
  \hline
  \hline
$ \frac{1}{21} $ &5.955(80)\\
$ \frac{1}{23} $ &5.908(61)\\
$ \frac{1}{25} $ &5.823(40)\\
$ \frac{1}{27} $ &5.698(26)\\
$ \frac{1}{29} $ &5.633(17)\\
$ \frac{1}{31} $ &5.628(23)\\
$ \frac{1}{33} $ &5.598(16)\\
$ \frac{1}{35} $ &5.574(13)\\
$ \frac{1}{37} $ &5.530(10)\\
$ \frac{1}{39} $ &5.514(11)\\
$ \frac{1}{41} $ &5.494(15)\\
$ \frac{1}{43} $ &5.497(12)\\
$ \frac{1}{47} $ &5.459(10)\\
$ \frac{1}{49} $ &5.460(8)\\
$ \frac{1}{57} $ &5.419(7)\\
$ \frac{1}{67} $ &5.392(6)\\
$ \frac{1}{79} $ &5.367(6)\\
  \hline
  \hline
\end{tabular}%
\caption{Negativity decay constants, $\beta_{2D}$, extracted at finite region pixelation appearing in the middle two panels of Fig.~\ref{fig:neg2d}.}
\label{tab:betaextrap}
\end{table}

\begin{table}[ht]
  \begin{tabular}{c|c}
  \hline
  \hline
  $d_{lat}$ & $\tilde{r}_{\Nslash}/d_{lat}$ \\
  \hline
  \hline
  1&1.000\\
2&1.000\\
3&3.000\\
4&3.250\\
5&5.200\\
6&5.333\\
7&7.429\\
8&7.625\\
9&9.667\\
10&9.800\\
11&11.91\\
12&12.00\\
13&14.15\\
14&14.21\\
15&16.33\\
16&16.50\\
\end{tabular}
\quad
\begin{tabular}{c|c}
17&18.59\\
18&18.72\\
19&20.79\\
20&20.95\\
21&23.05\\
22&23.18\\
23&25.26\\
24&25.42\\
25&27.52\\
26&27.65\\
27&29.74\\
28&29.86\\
29&32.00\\
30&32.10\\
31&34.23\\
32&34.34\\
33&36.45\\
\end{tabular}%
  \caption{Entanglement bubble radii for regions of the one-dimensional massless scalar field appearing in the right panel of Fig.~\ref{fig:neg2d}.}
  \label{tab:1dbubbleradii}
\end{table}

\begin{table}
  %
  \caption{Entanglement bubble radii for regions of the two-dimensional massless scalar field appearing in the right panel of Fig.~\ref{fig:neg2d}.}
  \label{tab:2dbubbleradii}
\end{table}

\end{document}